\documentclass{aa}  
\usepackage[varg]{txfonts}
\defcitealias{2004VOuchi}{O04}

\defcitealias{2006Yoshida}{Y06}
\usepackage{graphicx}
\usepackage[hidelinks,colorlinks=true,breaklinks,allcolors=blue]{hyperref} 
\usepackage{natbib}
\bibpunct{(}{)}{;}{a}{}{,} 
\begin{document} 
\title{An Overdensity  of Lyman Break Galaxies Around the Hot Dust-Obscured Galaxy WISE J224607.56$-$052634.9}

\titlerunning{An Overdensity  of LBGs Around W2246$-$0526}
\author{Dejene Zewdie
 \inst{1}\fnmsep\thanks{dejene.woldeyes@mail.udp.cl}
          \and
Roberto J. Assef\inst{1}
\and
Chiara Mazzucchelli  \inst{1}
\and
Manuel Aravena \inst{1}
\and
Andrew W. Blain \inst{2}
\and
Tanio D\'iaz-Santos\inst{3,4}
\and
Peter R. M. Eisenhardt\inst{5}
\and
Hyunsung D.~Jun\inst{6}
\and 
Daniel Stern\inst{5}
\and
Chao-Wei Tsai\inst{ 7,  8, 9}
\and
Jingwen W. Wu\inst{9, 7}
          }
\institute{Instituto de Estudios Astrof\'isicos, Facultad de Ingenier\'ia y Ciencias, Universidad Diego Portales, Av. Ej\'ercito Libertador 441, Santiago, Chile
         \and
Physics \& Astronomy, University of Leicester, 1 University Road, Leicester LE1 7RH, UK
    \and
Institute of Astrophysics, Foundation for Research and Technology–Hellas (FORTH), Heraklion, GR-70013, Greece
    \and 
School of Sciences, European University Cyprus, Diogenes street, Engomi, 1516 Nicosia, Cyprus
\and
Jet Propulsion Laboratory, California Institute of Technology, 4800 Oak Grove Drive, Pasadena, CA 91109, USA
    \and
SNU Astronomy Research Center (SNUARC), Astronomy Program, Department of Physics and Astronomy, Seoul National University, Seoul 08826, Republic of Korea
    \and 
    National Astronomical Observatories, Chinese Academy of Sciences, 20A Datun Road, Beijing 100101, China
    \and
    Institute for Frontiers in Astronomy and Astrophysics, Beijing Normal University, Beijing 102206, China    
    \and 
    University of Chinese Academy of Sciences, Beijing 100049, China   }             
   \date{Received: April 19, 2023; accepted: June 28, 2023}

  \abstract{ We report the identification of Lyman Break Galaxy (LBG) candidates around the most luminous Hot Dust-Obscured Galaxy (Hot DOG) known, WISE J224607.56$-$052634.9 (W2246$-$0526) at $z=4.601$, using deep \textit{r}-, \textit{i}-, and \textit{z}-band imaging from the Gemini Multi-Object Spectrograph South (GMOS-S). We use the surface density of LBGs to probe the Mpc-scale environment of W2246$-$0526 to characterize its richness and evolutionary state. We identify LBG candidates in the vicinity of W2246$-$0526 using the selection criteria developed by \cite{2004VOuchi} and \cite{2006Yoshida} in the Subaru Deep Field and in the Subaru XMM-Newton Deep Field, slightly modified to account for  the difference between the filters used, and we find 37 and 55 LBG candidates, respectively. Matching to the $z$-band depths of those studies, this corresponds to $\delta = 5.8^{+2.4}_{-1.9}$ times the surface density of LBGs expected in the field. Interestingly, the Hot DOG itself, as well as a confirmed neighbor, do not satisfy either LBG selection criteria, suggesting we may be missing a large number of companion galaxies. Our analysis shows that we are most likely only finding those with higher-than-average IGM optical depth or moderately high dust obscuration. The number density of LBG candidates is not concentrated around  W2246$-$0526, suggesting either an early evolutionary stage for the proto-cluster or that the Hot DOG may not be the most massive galaxy, or that the Hot DOG may be affecting the IGM transparency in its vicinity. The overdensity around W2246$-$0526 is comparable to overdensities found around other Hot DOGs and is somewhat higher than typically found for radio galaxies and luminous quasars  at a similar redshift. }


  \maketitle

\section{Introduction}
Active Galactic Nuclei (AGNs) are some of the most luminous non-transient objects in the Universe. Observational studies have found luminous AGN activity already at the epoch of reionization \citep[$z>5.5$; see][for a recent review]{2022Fan}, implying that very massive supermassive black holes (SMBHs) existed only a few hundred million years after the Big Bang  \citep[e.g.,][]{2017Mazzucchelli, 2021Yang, 2022Farina}. There is strong evidence that luminous AGN activity is linked to galaxy mergers \citep{2012Treister}, although the causality of this relation is still a matter of debate. Numerical simulations strongly suggest that in the early Universe, the most massive SMBHs reside in the densest regions, built up from the accretion and merger of massive dark matter halo seeds, and surrounded by a large number of fainter galaxies \citep{2005Springel, 2006Volonteri, 2014Costa, 2019Habouzit}. Mergers and high gas accretion rates, both associated with high-density regions, could be driving this very rapid growth, but the attempts to study the environments of these objects based on searches for Lyman-Break Galaxies \citep[LBGs; e.g.,][]{2003Steidel, 2006Zheng, 2007Kashikawa, 2010Utsumi,  2013Husband, 2014Morselli,   2017GarcaVergara} and Lyman Alpha Emitters  \citep[LAEs; e.g.,][]{2007Kashikawa, 2019Garcia} have produced a wide variety of results. This is particularly the case for $z\gtrsim  5$ luminous  quasars \citep[e.g.,][]{2009Kim, 2014Morselli}. Recently, multi-wavelength observational studies with the Very Large Telescope (VLT) and the Atacama Large Millimeter Array (ALMA), have found evidence for a strong clustering of LBGs, LAEs, and CO emitters around quasars at $z\sim 4$ \citep{2017GarcaVergara, 2019Garcia, 2022Garcia}. While searches with ALMA for [CII]158 $\mu$m line emitting companions around a few quasars at $z\sim 6$   \citep{2018Decarli}   and $z\sim 5$ \citep{2017Trakhtenbrot, 2020Nguyen}, have been successful, other studies that focused solely on continuum emitters/SMGs were not  \citep[e.g.,][]{2018Champagne, 2022Meyer}. These findings are supported by observations of a large number of star-forming galaxies highly clustered around quasars at $z\sim6$ \citep[e.g.,][]{2010Utsumi, 2017Decarli, 2020Mignoli}. Studies of radio-loud AGN at $z\sim2$ have found them to be good tracers of protoclusters \citep{2013Wylezalek, 2016Noirot,2017Magliocchetti, 2017Retana-Montenegro}. At high-redshift $(z \sim 5 - 6)$ a variety of results have been reported, with some authors finding these quasars trace overdensities \citep[][]{2006Ajiki, 2006Zheng, 2007Venemans, 2020Bosman}.  However, some studies found no significant excess of galaxies in high-redshift quasar fields \citep[e.g.,][]{2013Banados, 2014Simpson, 2017Mazzucchelli}, and other studies finding a mix of overdensities and under-densities in the vicinity of quasars at high-redshift \citep{2005Stiavelli, 2009Kim}.  

Hot dust-obscured galaxies \citep[Hot DOGs,][]{2012Eisenhardt, 2012Wu} are a population of hyper luminous, obscured quasars, which were identified using the Wide-field Infrared Survey Explorer \citep[WISE;] []{2010Wright}. Due to their high dust obscuration, these objects are detected by WISE at 12 and 22$\mu$m and are undetected or faint in the more sensitive 3.4 and 4.6$\mu$m bands. The extreme bolometric luminosities of Hot DOGs, $L_{\rm bol} > 10^{13} L_\odot$ (10\% of which exceed $10^{14} L_\odot, $\citealt{2015Tsai}), are powered by accretion onto SMBHs buried under enormous amounts of gas and dust, making them close to Compton-thick in the X-rays \citep{2014Stern, 2015Piconcelli, 2015Assef, 2016Assef, 2020Assef}. The obscuring material absorbs ultraviolet and optical light, re-emitting it as infrared light.  Hot DOGs exert significant feedback into their host galaxies by driving massive ionized gas outflows \citep{2016Tanio, 2020Finnerty, 2020Jun}.  Together, these studies suggest that Hot DOGs may be probing a critical stage in the evolution of their host galaxies, in which quasar feedback is starting to shut down star formation by ejecting significant amounts of cold gas and transitioning into a UV-bright traditional quasar \citep[e.g., see][]{2018Tsai, 2018Wu, 2020Assef}.   Notably, \cite{2015Assef} found that the number density of Hot DOGs is similar to that of equally luminous type 1 AGN at $2<z<4$, with an effective number surface density of 1 per $31\pm4$ deg$^2$.

Multiwavelength observations of Hot DOGs have statistically shown that these objects are likely to live in dense environments \citep{ 2014Jones, 2015Assef, 2019Penney}. More recently, \cite{2022Luo} found an overdensity of distant red galaxies (DRGs) around a Hot DOG at $z=2.3$, while \cite{2022Ginolfi} revealed an overdensity of LAEs in the environment of a $z=3.6$ Hot DOG with VLT/MUSE observations. All of this evidence suggests that Hot DOGs exist in overdense regions of the Universe, where significant gas accretion can be maintained and  mergers could lead to large-scale obscuration. Furthermore, \cite{2018Tanio} found that WISE J224607.56$-$052634.9 (hereafter, W2246$-$0526), the most luminous Hot DOG known with $L_{\rm Bol}=3.6  \times 10^{14} L_{\odot}$  \citep{2015Tsai} and at redshift $z=4.601$ \citep{ 2016Tanio}, is found at the center of a multiple merger systems \citep{2018Tanio}. Using deep ALMA observations, \cite{2016Tanio, 2018Tanio} revealed three spectroscopically confirmed companion galaxies with disturbed morphologies and four sources as potential companions within $\sim$30 kpc. They found that W2246-0546 lives in a dense local environment and speculate that it may become the brightest cluster galaxy (BCG) of a galaxy cluster at redshift 0. \cite{2016Tanio} showed that this object is likely in a key stage of its evolution, experiencing isotropic outflows of atomic gas. However, the evidence of overdense environments so far has come through deep ALMA observations encompassing a spatial scale of only $\sim$30 kpc. There have not yet been detailed studies of the environments of this object on larger scales ($>$30 kpc), relevant to assess its Mpc-scale environment. Further characterizing the environments of Hot DOGs, such as W2246$-$0526, is critical to understand how galaxies evolve over cosmic time and the physical processes driving their evolution.

The Lyman break identification technique \citep[e.g.,][]{1996Steidel, 1999Steidel, 2002Giavalisco} is commonly used to select star-forming galaxies at $z>2$.  Star-forming galaxies at high redshift selected with this method are known as LBGs. The technique uses broadband photometry, typically with three bands that bracket the Lyman break to identify sources that are faint in the bluest band due to redshifted hydrogen absorption but have the flat continuous UV emission typical of star-forming galaxies in the two redder bands. As the Lyman alpha forest opacity increases with redshift at $z>4$, a large number of LBGs have been identified using the combination of Lyman break and Lyman alpha absorption (\citealt{2004VOuchi}, hereafter \citetalias{2004VOuchi}; \citealt{2006Yoshida}, hereafter \citetalias{2006Yoshida}).

In this paper, we probe the environment within 1.4 Mpc of W2246$-$0526 by studying the surface density of nearby LBGs. Specifically, we study LBG candidates at redshift $ z \sim 4.601$ selected as \textit{r}-band dropouts in deep Gemini Multi-Object Spectrographs South (GMOS-S) imaging using the selection functions of \citetalias{2004VOuchi} and \citetalias{2006Yoshida}, modified to fit our specific set of filters.  The paper is organized as follows. In Section 2, we describe the observations from  GMOS-S, the data reduction, and the photometric measurements. In Section 3, we discuss modifications to the color selection function of \citetalias{2004VOuchi} and \citetalias{2006Yoshida} to accommodate the GMOS-S filters used. We also study the color and space distribution of the LBG candidates as well as their luminosity function. In Section 4, we compare our results with other Hot DOG and quasar environments in the literature. We present our conclusions in Section 5. Throughout this paper, all the magnitudes are in the AB system. We assume a flat $\Lambda$CDM cosmology with $H_0$ = 70 $\rm km~\rm s^{-1}~\rm Mpc^{-1}$ and $\Omega_{M}$= 0.3. At a redshift of z = 4.601, W2246$-$0526 is observed at 1.3 billion years after the Big Bang, and an observed spatial scale of 1$\arcsec$ is equal to 6.5 kpc.

 \section{Observations} \label{Sec2}
We obtained GMOS-S deep imaging in the \textit{r}-band on the nights of UT2017-09-16 and UT2017-09-23, in the \textit{i}-band on the nights of UT2017-09-16 and UT2017-09-27, and in the \textit{z}-band on the night of UT2017-09-23, (program ID: 102 207-33Q, P.I: R. J. Assef). During our observations, the average seeing conditions in the \textit{r}-, \textit{i}-, and \textit{z}-bands were 0.52$\arcsec$, 0.61$\arcsec$, and 0.58$\arcsec$, respectively. The observations were taken with an average airmass of  1.1 for the \textit{r}-band, 1.3 for the \textit{i}-band, and 1.2 for the \textit{z}-band. All images were obtained with a pixel scale of 0.16\arcsec~pix$^{-1}$. We obtained 25 and 24 images with exposure times of 300 seconds each in the \textit{r}- and \textit{i}-bands, respectively, and 62 in the \textit{z}-band with an exposure of 100 seconds to avoid the sky emission becoming non-linear. The observations are summarized in Table \ref{observation}.  The field of view of the instrument is 5.5 arcmin $\times$ 5.5 arcmin  and  observations were obtained by dithering with a grid of 4 by 3 with 6 arcsec intervals for the \textit{r}- and \textit{i}-bands, and a grid of 6 by 5 for the \textit{z}-band.

 \begin{table*} []
\caption[]{Information on the GMOS-S observations used in this work}
\label{observation}
 	$$ 
\begin{array}{c c c c c c cl}
	\hline
	\noalign{\smallskip}
 	\rm	Band  &  \text{\rm Exposure Time}   &  \text{\rm Mean Seeing}& \text{\rm Mean Airmass} & \text{\rm UT Dates}  & \text{Depth of the Stack Image}   \\
   &      &   &   &    &5 \sigma /3 \sigma /1 \sigma \\
  
 		\noalign{\smallskip}
 	\hline
 	\noalign{\smallskip}
 	r   & 25 \times 300s &  0.52\arcsec    & 1.1    & 2017-09-16/23 & 27.29/27.84/29.00\\
  i   & 24 \times 300s &  0.61\arcsec  &  1.3 &2017-09-16/27  & 26.58/27.16/28.34 \\
  z&62 \times 100s  &  0.58 \arcsec    & 1.2  &2017-09-23 & 26.02/26.57/27.77\\
 \noalign{\smallskip}
\hline
\end{array}
$$ 
 \end{table*}

   \begin{figure*}[ht]
 	\centering
 	\includegraphics[scale=0.56]{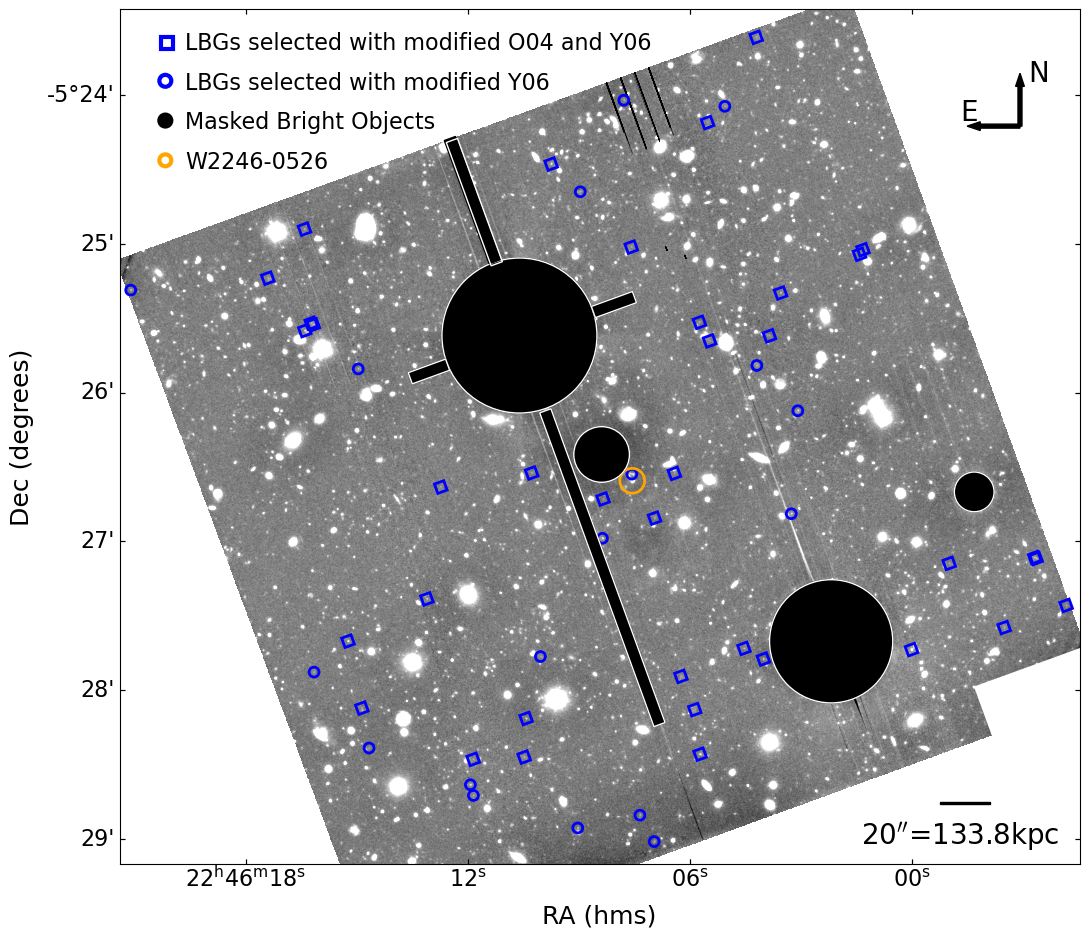} 
 	\caption{The \textit{i}-band image of W2246$-$0526. Black circles and rectangles show the masked area that we did not use for our LBG selection (see  Section \protect\ref{PhotometricData} for details). The 5 arcsec radius solid orange circle indicates the position of the W2246$-$0526, 4 by 4 arcsec blue rectangles show the LBG candidates selected with the modified \citetalias{2004VOuchi} and \citetalias{2006Yoshida} selection criteria  (see Table \ref{LBGO}),  and the 3 arcsec radius blue circles show those selected with the modified \citetalias{2006Yoshida} selection criteria only (see Figure \ref{colordiagram} and Table \ref{LBGY}). There are no LBG candidates present in the SE corner of the image that has been clipped.}
 	\label{iband}
 \end{figure*}

 \subsection{Data reduction} \label{reduce}
 
We reduced the GMOS-S data using the Gemini package in {\tt{IRAF}} V2.16 through the Python {\tt{PyRAF}}\footnote{\url{https://iraf-community.github.io/pyraf.html}} interface. We use the GMOS package to  bias-correct, and flatfield the images. We combined the images using the Gemini imcoadd task. We find, however, that the extended glow of two heavily saturated bright stars ($r < 20$ mag) in the field results in a very uneven background in the final image obtained using this task, so we have instead decided to subtract the extended glow of these stars from each frame before combining them.  Specifically, we masked all sources of each fully reduced frame detected at $>5\sigma$ by the  {\tt{DAOStarFinder}} routine of the {\tt{photutils}}\footnote{\url{https://photutils.readthedocs.io/en/stable/}} Python package. Most sources were masked with a 15-pixel radius circle, but larger radii were used for brighter sources. We modeled the source-subtracted images as a combination of two Moffat profiles (one for the extended glow of each star) and a flat background. Then, we subtracted the best-fit Moffat profiles and co-added the resulting images without correcting for dithering to create  an illumination correction frame. As observations for some bands were obtained across multiple nights, we created one illumination correction frame per night per band. Finally, each Moffat-profile subtracted frame was corrected by the  appropriate illumination frame\footnote{The scripts to subtract the Moffat-profiles, create and apply the illumination corrections can be found at \url{https://github.com/rjassef/GMOS_W2246_ICMS}}. The corrected frames were finally stacked using the imcoadd task of the Gemini {\tt{IRAF}} package.  The reduced \textit{i}-band image is shown in Figure \ref{iband}.

\subsection{Photometry}\label{PhotometricData}

Using the {\tt{astroalign}} Python module, we aligned the \textit{r}- and \textit{z}-band images to the \textit{i}-band image. We then used {\tt{SExtractor}}\footnote{ {\tt{SExtractor}} version 2.5.0 \url{https://www.astromatic.net/software/sextractor/}} \citep{1996Bertin} for source detection and photometry. Specifically, we measured the photometry in fixed 2$\arcsec$ diameter apertures with {\tt{SExtractor}} in dual image mode using the \textit{i}-band image for source detection. We used a detection and analysis threshold of at least 3 pixels detected above 1.5$\sigma$. We used a global model for the background with mesh and filter sizes of 32 and 3, respectively. We carried out the source detection in the \textit{i}-band and used those positions to obtain photometric measurements in all three bands. We masked out the bright stars and their  spikes and cut off the borders  of the images (as well as regions not covered by all three bands) to avoid spurious sources.   The final \textit{i}-band image with the masking applied is shown in Figure \ref{iband}. We estimated the usable area by generating $10^6$ random uniform points, distributed throughout the image, and counting the fraction of unmasked points. The usable area remaining after the masking is 23.7 arcmin$^2$. For reference, the masked regions in Figure \ref{iband} account for 13\% of the combined field of view.

The photometric calibration was carried out using data from the Panoramic Survey Telescope \& Rapid Response System  (Pan-STARRS) Survey \citep{2012Tonry}. We only considered point sources, selected using the probabilistic classification of unresolved point sources with {\tt{ps\_score}} greater than 0.83 as suggested by \cite{2018Tachibana}\footnote{\url{https://outerspace.stsci.edu/display/PANSTARRS/How+to+separate+stars+and+galaxies}}. The Pan-STARRS point sources were cross-matched  with the sources from our imaging using a 1 arcsec radius. We find 36 Pan-STARRS point sources within the unmasked area of our images. To estimate the photometric calibration constant, we only consider sources with magnitudes $19 < i < 21$ (21 sources), $19 < r < 21.7$ (18 sources), and $17 < z < 20.4$ (21 sources) in Pan-STARRS. These magnitude limits are meant to avoid sources bright enough to be in the non-linear regime of the GMOS observations as well as poorly detected sources in Pan-STARRS. We fit for the photometric calibration of the GMOS data in \textit{r}-band using $g-r$ and $r-i$ colors, in \textit{i}-band using $r-i$ and $i-z$ colors, and in \textit{z}-band using $i-z$ color. These corrections were required to obtain an accurate calibration (see Section \ref{modified}) given the differences between the instruments and filters. We applied the 3$\sigma$ detection catalog limit to the \textit{i} (27.16) and \textit{z} (26.57) bands.
 
 \subsection{Detection completeness}\label{complete}
 
\begin{figure}[ht]
 \includegraphics[scale=0.54]{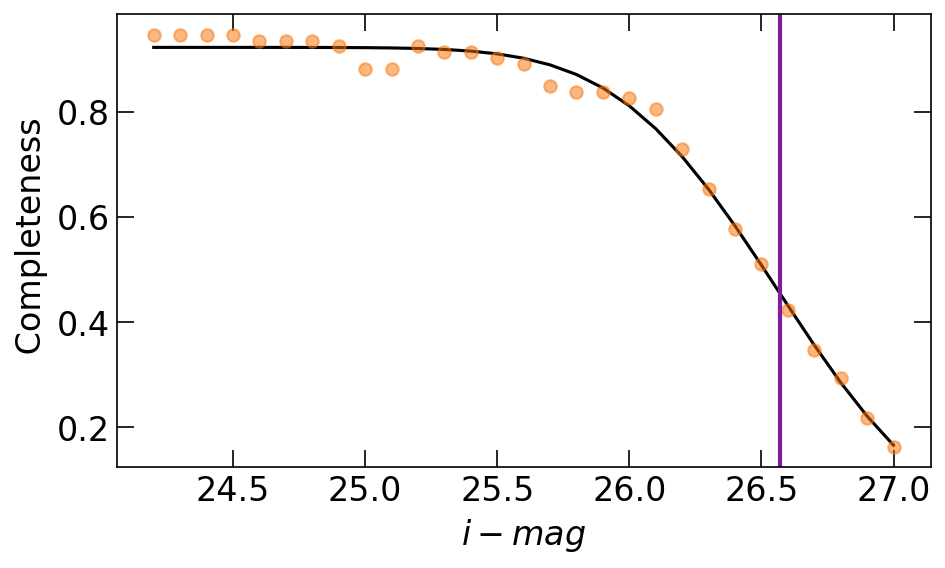}
 \caption{Detection completeness in the \textit{i}-band image. The dots show the completeness as a function of the apparent magnitudes in the \textit{i}-band solid  black line shows the best fit to the detection completeness. The vertical purple line shows the 3$\sigma$ limit of our \textit{z}-band observation.}  
 	\label{comp}
 \end{figure}

 We estimate the detection completeness as a function of the \textit{i}-band apparent magnitude since our LBG candidates were selected from sources detected in the \textit{i}-band image. The detection completeness was estimated using a Monte Carlo simulation. Specifically, we injected 100 mock sources into the \textit{i}-band image  with a Gaussian PSF with a full-width-at-half-maximum (FWHM) of 0.61 arcsecs (i.e., the average seeing in this band). We randomized the positions of the injected sources across the \textit{i}-band image and found that 92 sources fell within the unmasked region. We ran {\tt{SExtractor}} to recover the injected sources using the same procedure described in Section \ref{PhotometricData}. We repeated this process by changing the magnitude of the injected sources in the range between 24.1 to 27.0 mag in steps of 0.1 mag. The recovered fraction of injected sources is shown in Figure \ref{comp} as a function of the input magnitude. As expected, the completeness sharply decreases with increasing magnitude when approaching the 3$\sigma$ limit of the observations.

We fit the recovered fraction of injected sources using an error function.  Specifically, we fit the completeness $\rho$ as

 \begin{equation}
 	\rho (m_i)=\frac{1}{2}\alpha[1-erf(\frac{m_i-\mu}{\sigma})],
 \end{equation}
 \noindent where $m_i$ is the \textit{i}-band apparent magnitude, and $\alpha$, $\mu$, and $\sigma$ are fit to the data. We find that the completeness is best modeled by $\alpha$=0.92, $\mu$=26.56, and $\sigma$=0.68.

 \section{Selected Lyman Break Galaxy Candidates}\label{Sec3}
 
 \subsection{Modified method to identify LBG candidates}\label{modified}
 
\begin{figure*}[ht]
\includegraphics[scale=0.4]{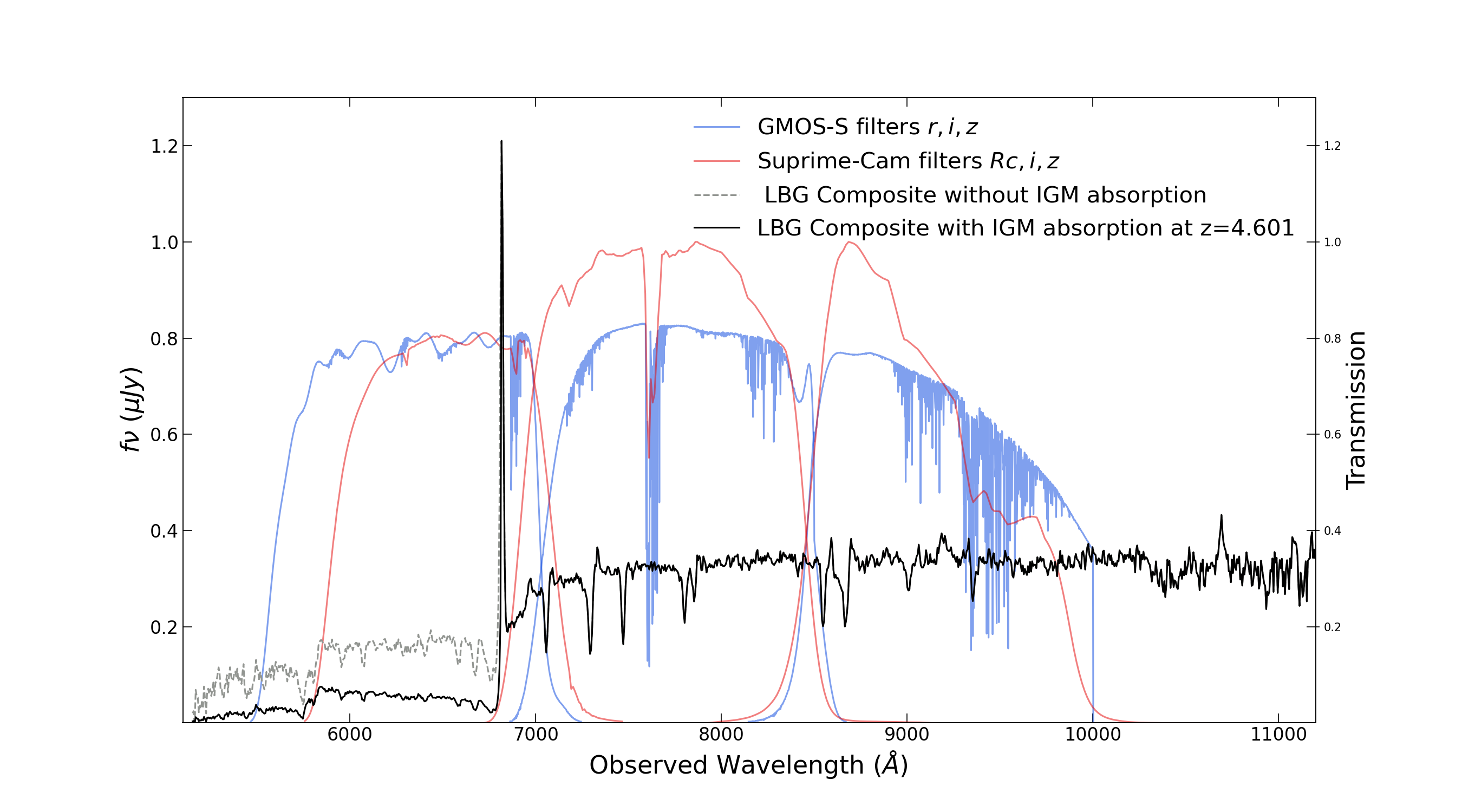}
\caption{Composite LBG spectrum of \protect\cite{2003Shapley}, with the IGM absorption at $z = 3$ and shifted to  $z=4.601$ (grey-shaded line), and the IGM absorption correction at $z = 4.6$ (black-solid line). See text for further details. After accounting for quantum efficiency and atmospheric transmission, the blue and the red solid line are the GMOS-S (used in this work) and Subaru Suprime-cam filter curves, respectively. }
\label{LBG_Spectrum}
 \end{figure*}

We used the LBG color selection criteria proposed by \citetalias{2004VOuchi} and \citetalias{2006Yoshida}. In  general, the \citetalias{2006Yoshida} criteria is a more permissive version of the \citetalias{2004VOuchi} criteria aimed at obtaining a larger number of LBG candidates. Both of them were developed using observations obtained with the Subaru Telescope  Suprime-cam instrument.  We consider them representative of the average surface density of sources in the sky. Both studies used the $R_{\rm c}$-, \textit{i}-, and \textit{z}-band imaging of the Subaru Deep Field (SDF) and while \citetalias{2004VOuchi} also used observations of the Subaru XMM-Newton Deep Field (SXDF). Since the Suprime-cam filters do not exactly match the GMOS-S \textit{r}-, \textit{i}-, and \textit{z}-bands (see Figure \ref{LBG_Spectrum}), we modify the selection function to ensure that the contamination level by interlopers of our sample is consistent with that of the samples of \citetalias{2004VOuchi} and \citetalias{2006Yoshida}. 
 
We used the $z\sim 3$  LBG composite spectrum from \cite{2003Shapley} to estimate the changes needed to be applied to the \citetalias{2004VOuchi} and \citetalias{2006Yoshida} selection criteria to accommodate the GMOS filter set. At z=4.601, we expect significantly more IGM absorption shortward of Ly$\alpha$  than at $z\sim 3$. Therefore, in order to incorporate the expected excess absorption at $z = 4.601$, we used the \cite{1995Madau} model, adding a parameter $\kappa$ to account for variations in the IGM optical depth. The equation we used to incorporate the dispersion and excess absorption is as follows:

 \begin{equation}\label{eq1}
 	f_{\nu}^{(\rm LBG) (z=4.6)} = f_{\nu}^{\rm (LBG) ( z=3)}  \rm exp(-\kappa \tau_{\rm eff (z=4.6)} + \tau_{\rm eff (z=3)}),
 \end{equation}

  \noindent where $\tau_{\rm eff}$ corresponds to the mean optical depth of the \cite{1995Madau} model.

 As can be seen in Figure 3 of \cite{1995Madau}, there is a significant dispersion around the mean optical depth of IGM absorption, $\tau_{\rm eff}$, at $z\sim 3.5$. We find that the 1$\sigma$ range in that Figure corresponds to $\kappa =$ 0.18 to 2.69 times the mean optical depth at $z\sim 3.5$. We later consider this range of $\kappa$ values when assessing the effectiveness of our selection.

Figure \ref{LBG_Spectrum} shows the LBG composite spectrum of \cite{2003Shapley}) shifted to $z=4.601$ assuming the mean IGM optical depth at this redshift (i.e., $\kappa=1$  see in eqn. [2]). The Figure also shows, for comparison, the same LBG composite spectrum but corrected for IGM absorption.

To estimate the modifications to the selection function,  we computed the expected photometry of this LBG template at $z=4.601$ for $\kappa=1$ and the Suprime-cam filters used by \citetalias{2004VOuchi} and \citetalias{2006Yoshida} and in the GMOS-S bands using the {\tt{synphot}} package\footnote{\url{https://synphot.readthedocs.io/en/latest/}.}. The transmission curves of the GMOS-S and Subaru Suprime-cam filters were obtained from the Spanish Virtual Observatory (SVO)\footnote{\url{http://svo2.cab.inta-csic.es/svo/theory/fps3/index.php?mode=browse}}. Subaru SuprimeCam's \textit{i} and \textit{z} filter transmissions are available considering the pass-band and atmospheric transmissions as well as the quantum efficiency, but for the $R_{\rm c}$ band only the passband transmission is available. For the GMOS-S filters, only the pass-band transmission is available. We used the quantum efficiency of the respective CCDs\footnote{quantum efficiency of Subaru Suprime-cam CCDs \url{https://www.subarutelescope.org/Observing/Instruments/SCam/ccd.html} and the GMOS-S CCDs \url{http://www.gemini.edu/instrumentation/gmos/components\#GSHam}} and the expected atmospheric transmission\footnote{Based on the location of the two instruments, we get the atmospheric transmission for both instruments using the Sky Model Calculator \url{https://www.eso.org/observing/etc/}} to construct the full transmission curves for these filters.

We calculated the colors of the LBG template redshifted  to $z=4.601$ with $\kappa=1$ to be  $(r-i)=1.18$, $(i-z)$=0.058, $(Rc-i_{\rm S})$=0.95 and $(i_{\rm S}-z_{\rm S})$=0.075. Applying this color difference, we can write the \citetalias{2004VOuchi}  selection criteria in  the GMOS-S bands as:

 \begin{multline}\label{eqO}
 	r-i>1.43, \\
	i-z<0.683,\\
 	r-i>i-z+1.247,\\
\end{multline}
 
 \noindent and the \citetalias{2006Yoshida} selection criteria in the GMOS-S bands as:
 
 \begin{multline}\label{eqY}
 	r-i>1.23,\\
 	i-z<0.683, \\
 	r-i> 1.2(i-z)+1.15. \\
 \end{multline}

 \begin{figure*}[ht]
 	\centering
 	\includegraphics[scale=0.7]{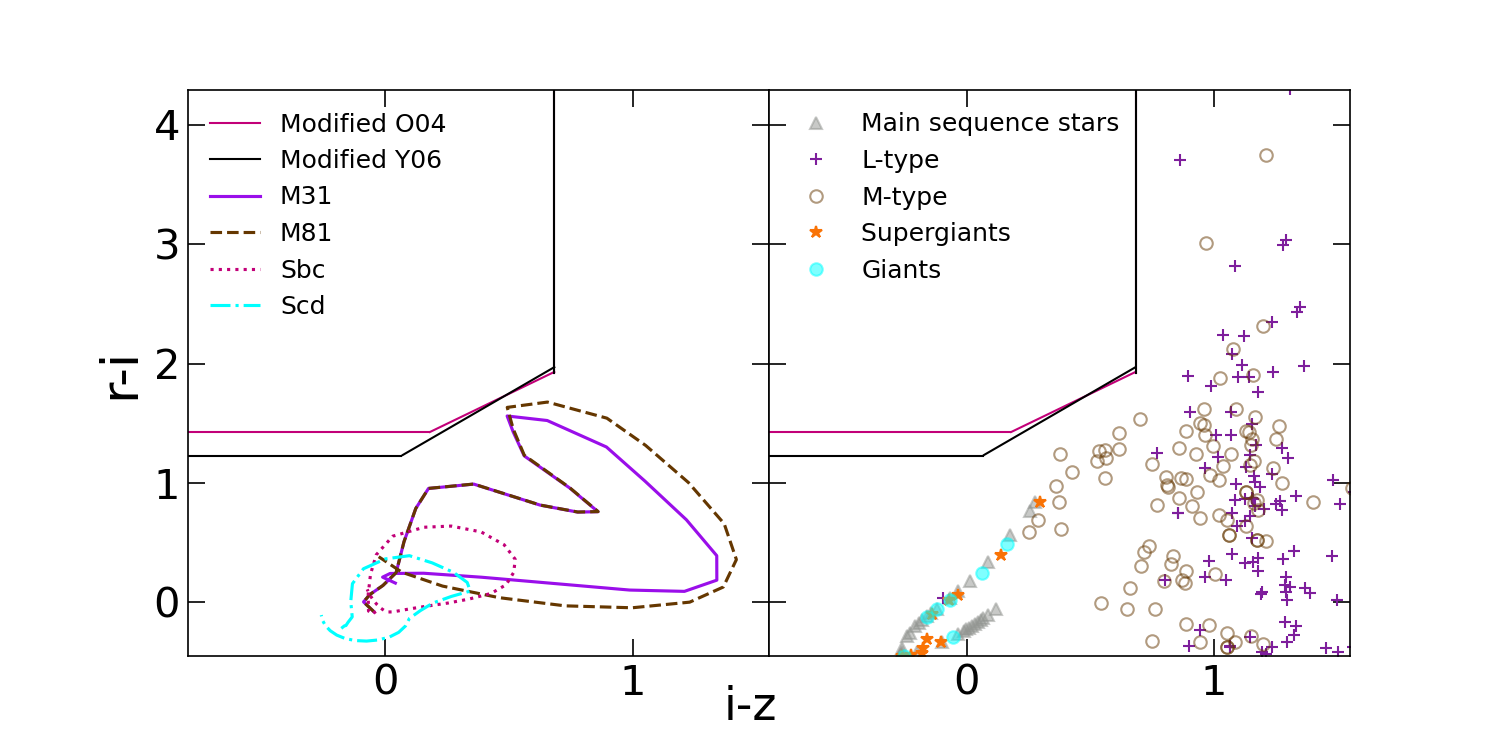}
 	\caption{ $r-i$ vs $i-z$ color-color diagram used in this work to select LBG candidates and exclude contaminants. The solid magenta and black lines define the selection boundaries in the color diagram established by the modified color selection criteria from \protect\citetalias{2004VOuchi}  and \protect\citetalias{2006Yoshida}, respectively. In the left panel, we show the representative colors of several classes of galaxies from \protect\cite{1980Coleman} as a function of redshift from z= 0 to 3. In the right panel we show the colors of different types of stars, from main-sequence to brown dwarfs, obtained by using models of stellar atmospheres from \protect\cite{2004Castelli} and  brown dwarf spectra from \cite{2014Burgasser}}. 
 	\label{inter}
 \end{figure*}

 To assess if our modified selection could be differently affected by contamination from stars and lower redshift galaxies, we follow a similar procedure as \citetalias{2004VOuchi}.  Figure \ref{inter} shows the colors of representative galaxy templates from \cite{1980Coleman}, shifted in the redshift range $0-3$. The figure also shows the colors of the main sequence, giant, and super-giant stellar atmosphere models of \cite{2004Castelli} and of the M and L dwarfs of \cite{2014Burgasser}. Figure \ref{inter} shows that our modified selection should not be more affected by contaminants than the \citetalias{2004VOuchi} selection (see their Figure 6) or the \citetalias{2006Yoshida} selection.

Figure \ref{colordiagram} shows the color tracks for the LBG composite spectrum of \cite{2003Shapley}, modified for the changes in the IGM absorption expected through equation (\ref{eq1}). We show the colors expected for the typical absorption ($\kappa=1$) as well as for $\kappa=0.18$ and $\kappa=2.69$ (see above). We find that neither the \citetalias{2004VOuchi} nor the \citetalias{2006Yoshida} selection is able to identify an LBG at $z=4.601$ with $\kappa=1$, assuming the \cite{2003Shapley} composite spectrum, as \citetalias{2004VOuchi} identifies sources with $z>4.715$, and \citetalias{2006Yoshida} with $z>4.625$. Figure \ref{colordiagram} also shows how colors at $z=4.601$ change due to dust absorption, assuming the reddening law of \cite{2000Calzetti}.  While the figure shows that LBGs with typical IGM absorption and no dust reddening would not be selected, an increase in either parameter can shift the sources into the selection region.   Figure \ref{reddein} shows the relation between the minimum value of $\kappa$ and the minimum value of $E(B-V)$ needed to shift our assumed LBG spectrum (eqn. [\ref{eq1}]) into the \citetalias{2004VOuchi} and  \citetalias{2006Yoshida} selection boxes. While unreddened sources would need $\kappa=1.4$ ($\kappa=1.08$) to appear inside the \citetalias{2004VOuchi} (\citetalias{2006Yoshida}) selection box, sources with $\kappa=1$ would only need a moderate amount of obscuration with $E(B-V)=0.34$ [$E(B-V)=0.08$] to be shifted into the selection region. We note too that as reddening affects the $i-z$ color, a maximum amount of reddening of $E(B-V)=1.53$ can be tolerated before $z=4.601$ LBGs escape through the left boundary of the selection region. This suggests our selection is likely missing a significant amount of sources with less reddening/IGM absorption than the minima in Figure \ref{reddein}.

Although we assume all sources are at $z=4.601$ (in contrast to the assumption of $z = 4.9$ by both \citetalias{2004VOuchi} and \citetalias{2006Yoshida}), we need to find out the redshift range over which our modified criteria can find LBGs. As shown in Figure \ref{colordiagram}, the modified selection criteria of \citetalias{2004VOuchi} yield redshift intervals of $4.715<z<5.361$ and of \citetalias{2006Yoshida} are  $4.625<z<5.361$, for $\kappa=1.0$, the mean IGM absorption of \cite{1995Madau}. While we assume $\kappa$=1 for simplicity to estimate the volume over which our sources are found, for completeness we note that the redshift range for $\kappa=2.69$, the upper 1$\sigma$ range of the \cite{1995Madau} IGM absorption distribution, is $z=4.477$ to $z=5.277$ which is comparable to the range found by \citetalias{2004VOuchi} ($z=4.6-5.2$) and when we used the modified selection criteria of \citetalias{2006Yoshida} ($z=4.7 - 5.1$), we found the redshift range to be $z=4.401$ to $z=5.277$. 
 
We tested the modified selection function that would be obtained by doing the same analysis but using instead the mock spectrum at $z = 4.601$ from the {\tt{JWST}} Advanced Deep Extragalactic Survey (JADES; \citealt{2018Williams}). We find that using these spectra instead of the \cite{2003Shapley} composite to calculate the color changes between the filter sets results in a very similar, yet slightly more permissive version of the selection criteria in equations  (\ref{eqO}) and (\ref{eqY}).  Given this, and the fact that the JADES mock spectrum is estimated for much fainter galaxies than the LBG candidates targeted here, we use the selection based on the \cite{2003Shapley} LBG composite spectrum.

 \subsection{Selected LBG candidates}\label{LBGSelection}
 
 Using the modified \citetalias{2004VOuchi} selection criteria, we identified 41 LBG candidates within the unmasked regions of our images. Of these, 4 have their photometry potentially contaminated by nearby CCD features as assessed by visual inspection, so we discard them. Of the remaining 37 LBG candidates, 27 are brighter than the 1$\sigma$ limit in the \textit{r}-band.  Using the modified \citetalias{2006Yoshida} selection criteria, we identified instead 55 candidates unaffected by CCD features and 45 of them are brighter than the 1$\sigma$ limit in the \textit{r}-band. Photometry and cutout images in \textit{r}, \textit{i}, and \textit{z}-band of all candidates and the Hot DOG provided in the Appendix in Tables \ref{LBGO} and \ref{LBGY} and Figures \ref{stamps} - \ref{stampy}.

 \begin{figure*}[ht]
 \includegraphics[scale=0.65]{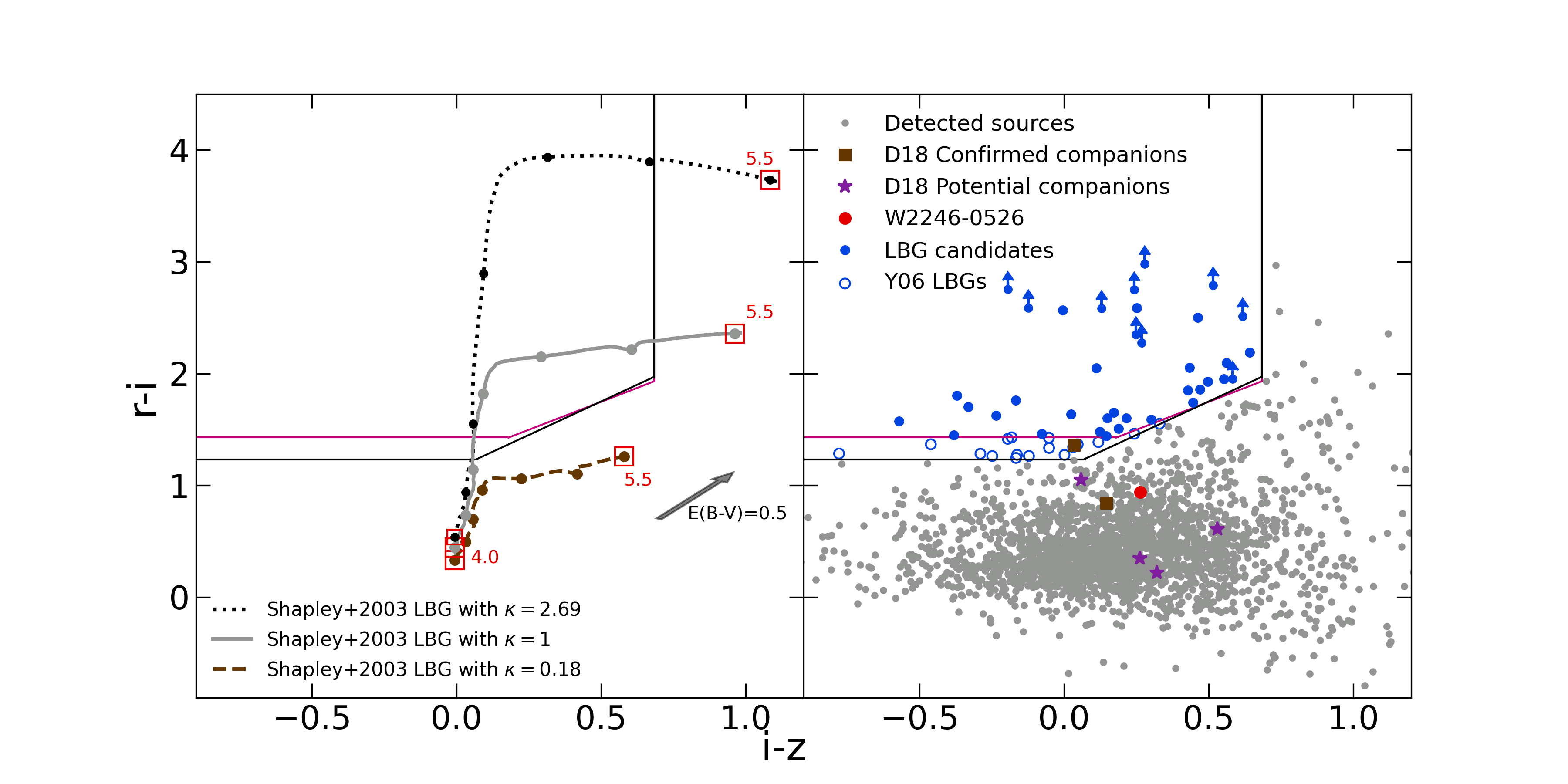}
 \caption{The left panel: shows the color-redshift track of the LBG composite spectrum of \cite{2003Shapley} with the IGM absorption of \cite{1995Madau} as described in eqn. (\ref{eq1}) for values of $\kappa$= 2.69 (dotted black line), 1.0 (solid gray line), and 0.18 (dashed brown line). The open red squares mark the $z=4.0$ and $z=5.5$ endpoints of each color-redshift track.  The dots in each track indicates $\Delta z =0.25$ bins.  The right panel shows the distribution of the $r-i$ and $i-z$ colors of sources around  W2246$-$0526 (gray dots). The blue-filled circles show the LBG candidates identified by both the \citetalias{2004VOuchi} and \citetalias{2006Yoshida} selection criteria.  The blue open circles show the LBG candidates identified by only the \citetalias{2006Yoshida} selection criteria.  The spectroscopically confirmed (brown squares) and potential (purple star) companions from \cite{2018Tanio} are also shown, and the red dot shows the Hot DOG.  Ten of the identified LBG candidates are fainter than 1$\sigma$ in the \textit{r}-band, and hence their ($r-i$) Colors are shown as lower limits.} 
 	\label{colordiagram}
 \end{figure*}

\begin{figure}[ht]
\includegraphics[scale=0.45]{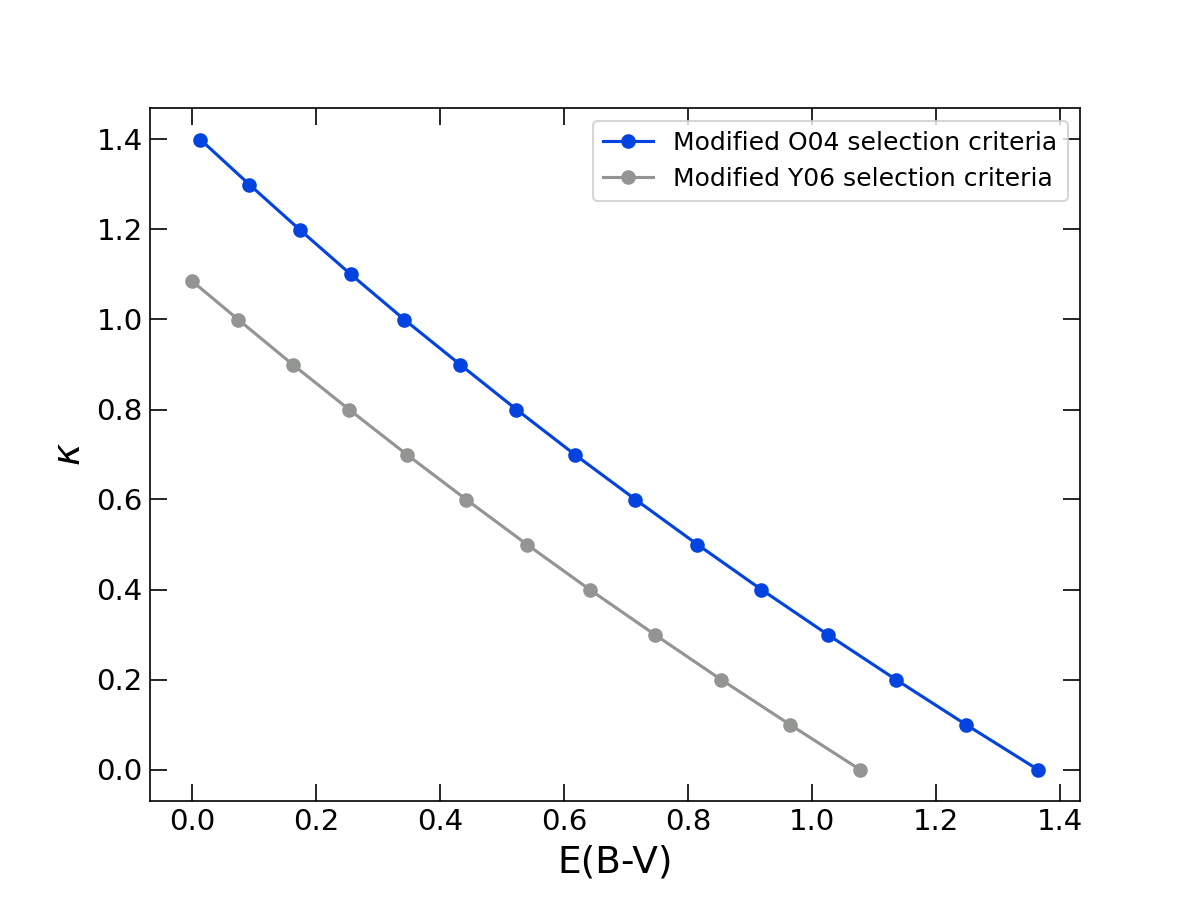}
\caption{The relation between the minimum value of the $\kappa$ and the minimum value of $E (B-V)$ that shift the \cite{2003Shapley} LBG composite spectrum spectrum at $z=4.601$ into the modified selection criteria of \citetalias{2004VOuchi} and \citetalias{2006Yoshida}. In both modified selection criteria, the selected sources with $\kappa=1$ would only require a moderate amount of obscuration with $E(B-V)=0.34$ and ($E(B-V)=0.08$) to be shifted into the modified \citetalias{2004VOuchi} and \citetalias{2006Yoshida} selection box} 
\label{reddein}
 \end{figure}

\begin{figure*}[ht]
\begin{center}
\includegraphics[scale=0.50]{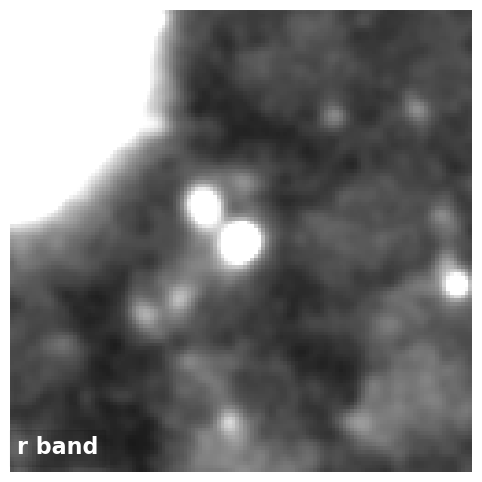} \hspace{0.1cm} \includegraphics[scale=0.5]{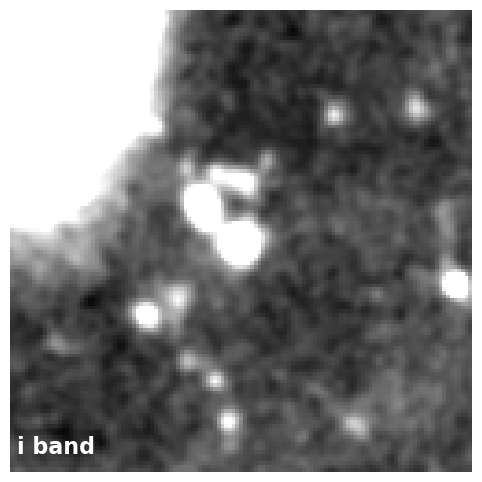}\\ 
\includegraphics[scale=0.5]{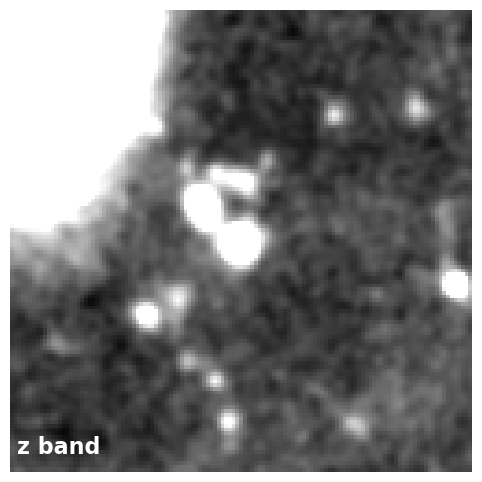}  \hspace{0.1cm} \includegraphics[scale=0.54]{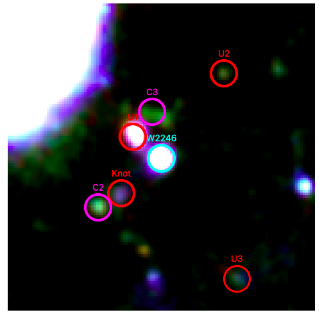}
\end{center}
\caption{Postage stamps (20$\arcsec \times 20\arcsec$) of W2246-0526 with similar field of view of deep ALMA observation by \cite{2016Tanio} and \cite{2018Tanio}. The top-left panel is \textit{r}-band, the top-right panel is the \textit{i}-band, the bottom-left panel is \textit{z}-band and the bottom-right panel is the color composite of W2246-0526. The spectroscopically confirmed and potential companions of \cite{2016Tanio,2018Tanio} are  shown in magenta and red circles, respectively. The Hot DOG is shown by the cyan circle. Cutouts of 8$\arcsec \times 8\arcsec$ of each of the marked sources are shown in Figure \ref{stampt}.} 
\label{stampw}
\end{figure*}

Figure \ref{colordiagram} shows the color distribution  diagram of all detected objects in the W2246$-$0526 field, with those selected as LBG candidates highlighted in blue.  We also highlight the locations of W2246$-$0526  and of the companions (both potential and with spectroscopic confirmation) identified with ALMA data by \cite{2016Tanio} and \cite{2018Tanio}. Interestingly, the modified \citetalias{2004VOuchi} selection criteria do not identify the Hot DOG, and the previously identified companions \citep{2018Tanio} as LBG candidates. However, the spectroscopically confirmed companions are closer to the selection box than the potential companions, suggesting the latter may be interlopers. Note that we are comparing populations of detected galaxies that were obtained in different ways. Using the modified \citetalias{2006Yoshida} selection criteria, we were able to identify one of the spectroscopically confirmed companions as an LBG candidate. Figure \ref{stampw} shows a zoomed-in view of a field that closely resembles the field of view of the deep ALMA observations of \cite{2018Tanio}. Additionally, we have marked the locations of spectroscopically confirmed and potential companion galaxies identified by \cite{2016Tanio} and \cite{2018Tanio}, and their stamps of 8$\arcsec\times 8\arcsec$ are shown in Figures  \ref{stampt}.

\begin{figure*}[ht]
 	\centering
 	\includegraphics[scale=0.75]{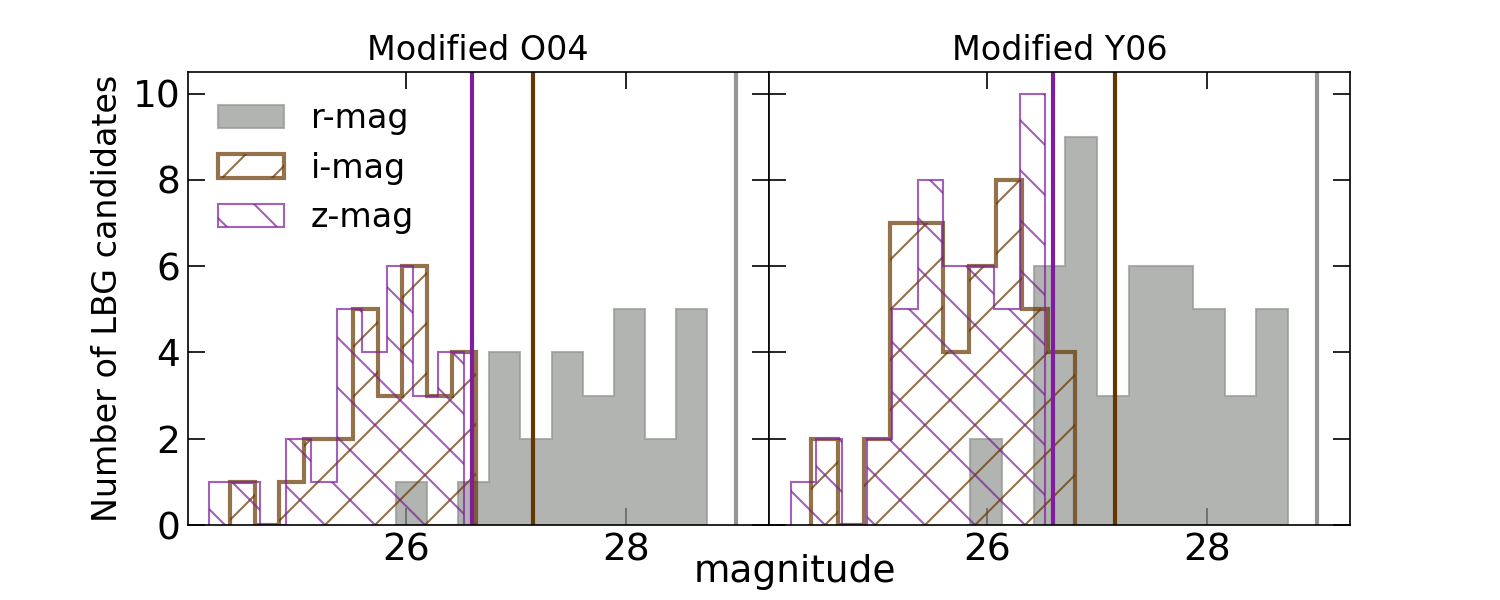}
 	\caption{The left and right panels show the magnitude distribution of the LBG candidates selected with the modified \citetalias{2004VOuchi} and \citetalias{2006Yoshida} selection criteria, respectively, in each band. The \textit{r}-band magnitude distributions only show the LBG candidates that are brighter than the 1$\sigma$ limit. The three vertical lines are the 3$\sigma$ magnitude limit for the \textit{i}-band (brown) and \textit{z}-band (purple) and the 1$\sigma$ magnitude limit for the \textit{r}-band (gray).}
 	\label{mag}
 \end{figure*}

Figure \ref{mag} shows the magnitude distribution of the selected LBG candidates in the three bands. For the \textit{r}-band, we only show sources brighter than the 1$\sigma$ limit.   The \textit{i}- and \textit{z}-band magnitudes have similar distributions, which is expected since LBGs have, by definition, flat UV SEDs. Figure \ref{iband} shows the \textit{i}-band image with the position of  the selected LBG candidates based on the modified \citetalias{2004VOuchi} and \citetalias{2006Yoshida} selection criteria.

 \subsection{Surface density of selected LBG candidates}\label{SD}
 
\begin{figure*}[ht]
\begin{center}
\includegraphics[scale=0.55]{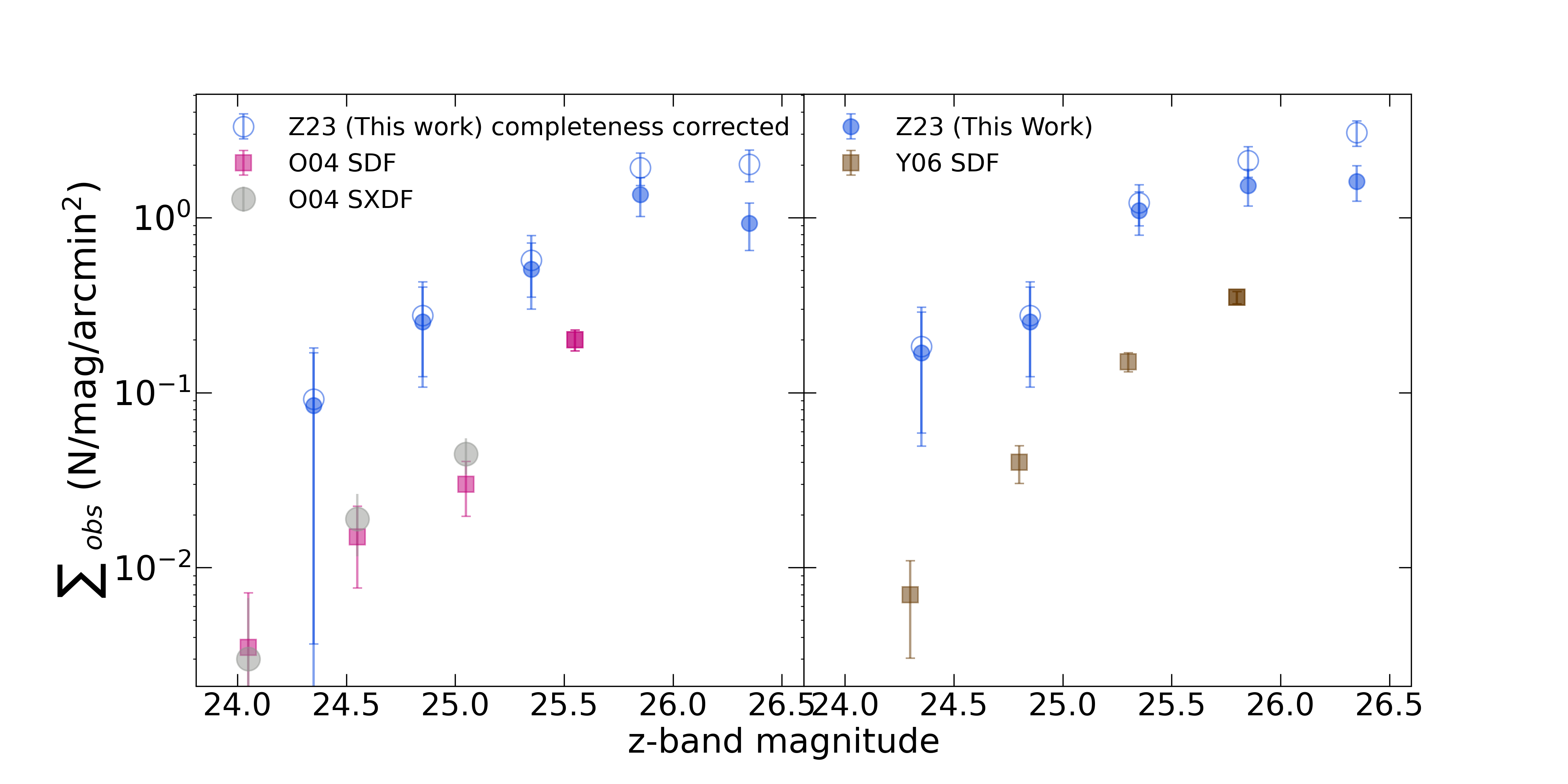} 
\end{center}
\caption{The surface density of the selected LBG candidates  (filled blue circles) and  LBG candidates corrected for completeness (open blue circles) compared to that measured in the SDF field by \citetalias{2004VOuchi} and \citetalias{2006Yoshida} (magenta and brown squares, respectively). We also show the measurement of LBG number counts in the SXDF field by \citetalias{2004VOuchi} (filled grey circles).} 
\label{Surface}
\end{figure*}

Figure \ref{Surface} compares the number counts of LBG candidates in the field of W2246$-$0526, as a function of apparent \textit{z}-band magnitude, to those measured by \citetalias{2004VOuchi} in the SDF and SXDF areas, and to those measured by \citetalias{2006Yoshida} using only the  SDF. In both studies, the number counts were not corrected for detection completeness. Figure \ref{comp} shows the detection completeness corrections and the corrected number counts in the field of W2246$-$0526 are shown alongside the uncorrected ones in Figure \ref{Surface}. For simplicity, we estimate the uncertainties of our surface density as the square root of the number of LBG candidates in each magnitude bin, neglecting cosmic variance.

Based on the modified \citetalias{2004VOuchi} and \citetalias{2006Yoshida} selection criteria, we find 37 and 55 LBG candidates, respectively, by matching the \textit{z}-band magnitude depth of in the field studies. We estimate the overdensity, $\delta =N_{found}/N_{expected}$, to be $\delta = 7.1\pm 1.1$ ($\delta = 5.1 \pm 1.2$) when compared with the \citetalias{2004VOuchi} study of the SDF (SXDF). On the other hand, we measure $\delta =5.2\pm 1.4$ when compared with the \citetalias{2006Yoshida} study of the SDF field. The average overdensity is hence $\delta = 5.8^{+2.4}_{-1.9}$.

 \subsection{Spatial distribution of the LBG candidates}\label{ND}
\begin{figure*}[ht]
 	\begin{center}
 		\includegraphics[scale=0.55]{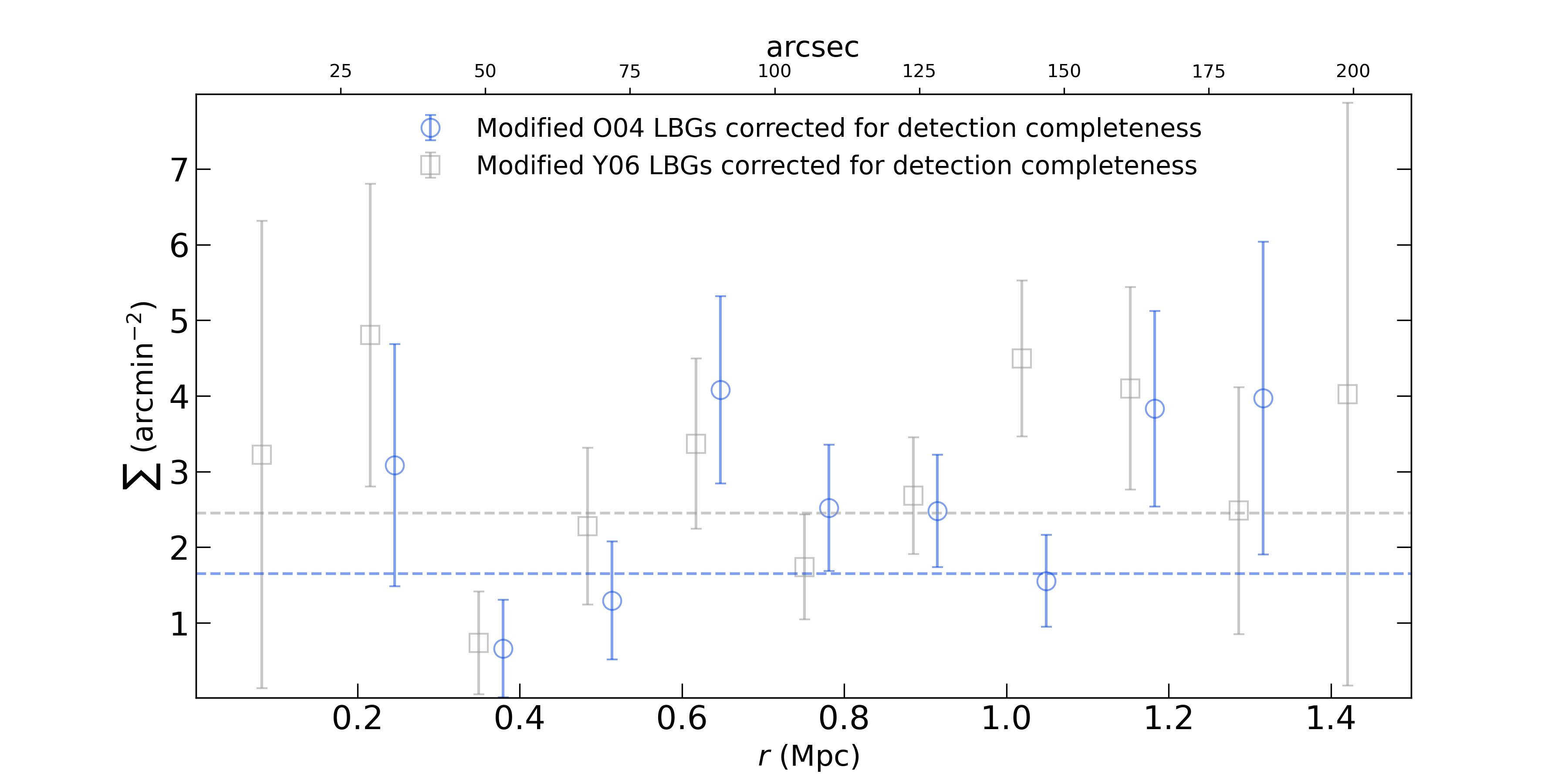}  
 	\end{center}
\caption{ The spatial distribution of LBG candidates as a function of distance to W2246$-$0526. The blue open circles and gray open squares show the selected LBG candidates based on the modified \citetalias{2004VOuchi} and \citetalias{2006Yoshida} selection criteria, respectively, corrected for detection completeness. We count the number of LBG candidates in annuli with 20 arcsec radius intervals avoiding the inner 2 arcseconds ($\sim$13 kpc).  For clarity, we shift the surface density of the \citetalias{2004VOuchi} selected LBG candidates by $+0.03$ Mpc on the x-axis.}
 	\label{distance}
 \end{figure*}

To gauge whether the two-dimensional overdensity is concentrated around W2246$-$0526, we count the number of LBG candidates in annuli with widths of 20 arcsecs, centered on W2246$-$0526 and starting 2 arcsecs ($\sim$13 kpc) away from the Hot DOG, The brightness of W2246-0526 interferes with the detection of candidates at smaller radii. We estimate the unmasked area of each ring using randomly distributed points within each ring and counting the unmasked fraction.  Figure \ref{distance} shows the density of LBG candidates as a function of distance to the Hot DOG.  While the highest density of LBG candidates is in the second ring around W2246$-$0526, there is not a clear radial profile centered on this Hot DOG. As done earlier, for simplicity, we estimate the uncertainties as the square root of the number of LBG candidates found in the ring divided by the usable area of the ring.

 \subsection{Luminosity function of Lyman break galaxy candidates}
 
We estimate the UV luminosity function of our LBG candidates at rest-frame wavelength 1700\AA\ following the approach of \citetalias{2004VOuchi}. Specifically, we estimate a correction to transform the\textit{ z}-band (rest-frame effective wavelength = 9521.7\AA\ at $z=4.601$) magnitude into a monochromatic flux at 1700\AA\ using the {\tt{synphot}} package with the LBG composite of \cite{2003Shapley}. We assume a top-hat band pass at rest-frame 1700\AA\ with a  $\pm 5$\AA\ width. Given that the \textit{z}-band is already covering a region close to that of rest-frame 1700\AA, we find an almost negligible $k$ correction of 0.022 mag. Finally, we convert the \textit{z}-band apparent magnitude by using the following equation.

 \begin{equation}
 	M_{1700 \textup{~\AA}}=m_z+2.5\log(1+z)-5\log dl+5 -k ,
 \end{equation}
 
 \noindent where $m_z$ is the apparent magnitude in the \textit{z}-band and $dl=42305.5$ Mpc, is the luminosity distance at $z=4.601$.
 
 Similarly,  we follow \citetalias{2006Yoshida} to estimate the UV luminosity function of our LBG candidates at 1500\AA\ in the rest-frame for which we calculate a $k$ correction of -0.015.

 We estimate the effective volume $V_{\rm eff}$ on which our sample is found as the comoving volume between the $z_{\rm min}$ and $z_{\rm max}$ redshifts of each selection (see Section \ref{complete}) within our field of view of 23.7\arcmin$^2$. For the \citetalias{2004VOuchi} and \citetalias{2006Yoshida} selections, this corresponds to 20638.77~Mpc$^{-3}$ and 23633.09~Mpc$^{-3}$ respectively. We computed the luminosity function from the observed number density of the completeness corrected \textit{z}-band at the UV rest-frame using these comoving volumes:

 \begin{figure*}[ht]
 	\begin{center}
 		\includegraphics[scale=0.5]{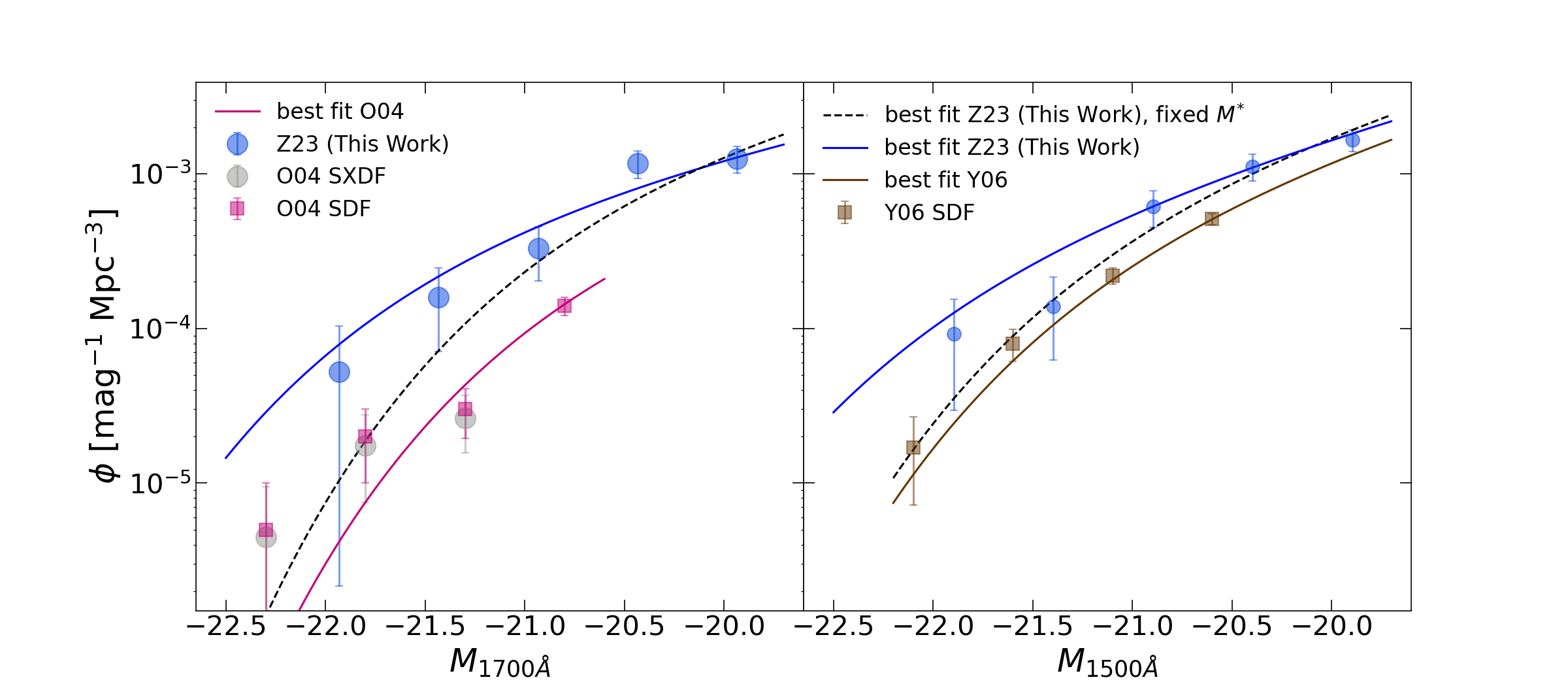}
 	\end{center}
 	\caption{The UV luminosity functions of the LBGs at $z\sim 4.6$. The blue circles are selected LBG candidates around the Hot DOGs. The magenta and brown rectangles are from \citetalias{2004VOuchi} and \citetalias{2006Yoshida},  respectively. The solids line shows the best-fit Schechter function for each study and the black dashed lines show the best fit by fixing the $M_{*}$ and $\alpha$ based on the result of \citetalias{2004VOuchi} and \citetalias{2006Yoshida}. }
 	\label{LF}
 \end{figure*}

 \begin{equation}
 	\phi(m)= \frac{1}{ V_{\rm eff} \Delta m} \sum_j \frac{1}{\rho(m_i)},
 \end{equation}
 
 \noindent where $\rho(m_i)$ is the detection completeness based on the \textit{ i}-band magnitude of each source j (see Section \ref{complete}). Figure \ref{LF} shows the estimated luminosity functions.
 
   We fit the luminosity function using a \cite{1976Schechter} function, namely,

 \begin{multline}
 	\phi (M_{\rm UV})dM_{\rm UV}=\left(0.4\ln 10\right)\phi^{*}\left( 10^{0.4(M_{\rm UV}^{*}-M_{\rm UV})}\right)^{\alpha +1}\\ \exp \left(-10^{0.4(M_{\rm UV}^{*}-M_{\rm UV})} \right)dM_{\rm UV},
 \end{multline}
 
 \noindent where $M^{*}_{\rm UV}$ is the characteristic magnitude, $\phi ^{*} (h_{70}^{3}Mpc^{-3})$ is the normalization that provides the number density and $\alpha$ is the slope of the luminosity function at the faint end.

  As can be seen in Figure \ref{LF} the luminosity function of the LBG candidates around W2246$-$0526 has significantly larger space density than in the SDF studies of \citetalias{2004VOuchi} and \citetalias{2006Yoshida}, as expected given our results based on the number densities (see Figure \ref{Surface}). The  best-fit parameters of the luminosity function are shown in Table \ref{LumF}, for the modified \citetalias{2004VOuchi} and \citetalias{2006Yoshida} selection functions.

  \begin{table*}[ht] 
   \begin{center}
       	\caption[]{Parameters of the Luminosity Function calculated in this work, using the modified \protect\citetalias{2004VOuchi} and \protect\citetalias{2006Yoshida} criteria for selecting LBG candidates.}
 	\label{LumF}
 	$$ 
 	\begin{array}{c c c c c c c|}
 		\hline
 		\noalign{\smallskip}
 			& {\rm O04} &     &  & {\rm Y06} & \\
 				\phi^{*} (h_{70}^{3}Mpc^{-3})& M^{*}_{1700}(mag) &\alpha & \phi^{*} (h_{70}^{3}Mpc^{-3})& M^{*}_{1500}(mag) &\alpha\\
 		\noalign{\smallskip}
 		\hline
 		\noalign{\smallskip} 
 	2.487\times 10^{-3} & -20.3^{\dagger}& -1.6^{\dagger}&  1.78\times 10^{-3} & -20.72^{\ddagger}& -1.82^{\ddagger}\\  
 	9.557 \times 10^{-4} & -21.18\pm0.70& -1.6^{\dagger}& 7.029 \times 10^{-4} & -21.55\pm0.44& -1.82^{\ddagger}\\
 		\noalign{\smallskip}
 		\hline
 	\end{array}
 	$$ 
   \footnotesize $\dagger$: Value fixed to that used by \protect\citetalias{2004VOuchi}.\\
   \footnotesize $\ddagger$: Value fixed to that used by \protect\citetalias{2006Yoshida}. 
   \end{center}
 \end{table*}

 \section{Discussion} \label{sec:Discussion}

  \subsection{Overdensities Around  Hot  DOGs}
  In Section \ref{SD}, we showed that LBGs are more abundant by a factor of $\sim 6$ around W2246$-$0526 than in a blank field, leading to the conclusion that W2246$-$0526 lives in a dense region at a time when the Universe was 1.3 Gyr old. This result is qualitatively consistent with the conclusions of other studies. In particular, \cite{2015Assef} studied the environment of a large number of Hot DOGs using \textit{Spitzer}/IRAC imaging. Based on the number counts of red galaxies within 1$\arcmin$ arcminute of the Hot DOGs, \cite{2015Assef} found that, statistically, these objects live in dense environments, comparable to those of radio-loud AGN found by \cite{2013Wylezalek}.  Similarly, \cite{2014Jones, 2017Jones} concluded that  Hot DOGs live in overdense environments based on the excess of companions detected at sub-mm wavelengths with JCMT/SCUBA-2.

 More directly, \cite{2016Tanio, 2018Tanio}  studied W2246$-$0526 using deep ALMA observations. They found three spectroscopically confirmed companions joined by dusty streamers within $\sim$35 kpc of W2246$-$0526.  Additionally, they found 4 more continuum sources  in a $\sim 20\arcsec$ field of view that may correspond to further companions.  \cite{2018Tanio} argue that this Hot DOG is at the center of a triple merger and speculate its environment may become a massive cluster at $z=0$ with the Hot DOG as the BCG. However, we do not find a strong indication that the Hot DOG is at the center of the overdensity. This may be because a) the overdensity, most likely a proto-cluster, is in a very early stage of collapse and is far from being virialized; b) this Hot DOG is not the most massive, central galaxy of the structure, even if it is the most luminous and radiating well above its Eddington limit \citep{2018Tsai}; or c) the Hot DOG may be luminous enough to affect the IGM transparency around it, effectively erasing the LBG overdensity's radial pattern.   We notice that option b) appears unlikely given the lack of other bright sources in the field. However, {\tt{JWST}} observations probing the peak of stellar emission are needed to confirm this. We note that possibility c) is consistent with the fact that the spectroscopically confirmed companions of W2246$-$0526, as well as W2246$-$0526 itself, are not identified as LBG candidates. The potential consequence of the Hot DOG's feedback on the IGM is that we may be underestimating the number of companions in the vicinity of W2246-0526. 

 

Recently,  \cite{2022Ginolfi} studied the environment of W0410-0913, a Hot DOG at z = 3.631, using VLT/MUSE observations,  and also concluded that this Hot DOG lives in a dense environment. Specifically, they found the number density of Ly$\alpha$ emitters within 0.4 Mpc of W0410-0913 to be $14_{-8}^{+16}$ times that of field galaxies with comparable Ly$\alpha$ luminosity. Also recently, \cite{2022Luo} studied the environment of W1835+4355 at  $z=2.3$ and found a factor of two enhancement in the number of DRGs in its surroundings. The overdensity differences between these studies  may be indicative of the diversity of the environments around Hot DOGs, but may also be driven by the specific selection methods used in each study.  Cosmological simulations show that the presence of early SMBHs with masses of $\sim 10^9 M_{\odot}$ at $z\geq 6$ is typically associated with galaxy overdensities, although the number of galaxies involved can vary \citep[e.g., ][]{2014Costa, 2019Habouzit}. \cite{2019Habouzit} conducted a study using the Horizon AGN simulations, which focus on a wide range of galaxy and black hole properties, as well as the potential overdensity of the regions. They predict a wide variety of both obscured and unobscured quasar environments that are not driven by cosmic variance.  While the use of photometric selection presents challenges due to cosmic variance, the inclusion of a large sample and the consideration of spectroscopic confirmation has the potential to provide additional insights \citep{2023Champagne, 2023Wang}. Despite using a photometric selection and focusing on a single Hot DOG environment, the observed level of overdensity of LBG candidates around the Hot DOG could potentially represent a lower limit if spectroscopically confirmed.

 \subsection{Comparisons Overdensities Around  Other Luminous Sources}
\begin{figure*}[ht]
\begin{center}
\includegraphics[scale=0.6]{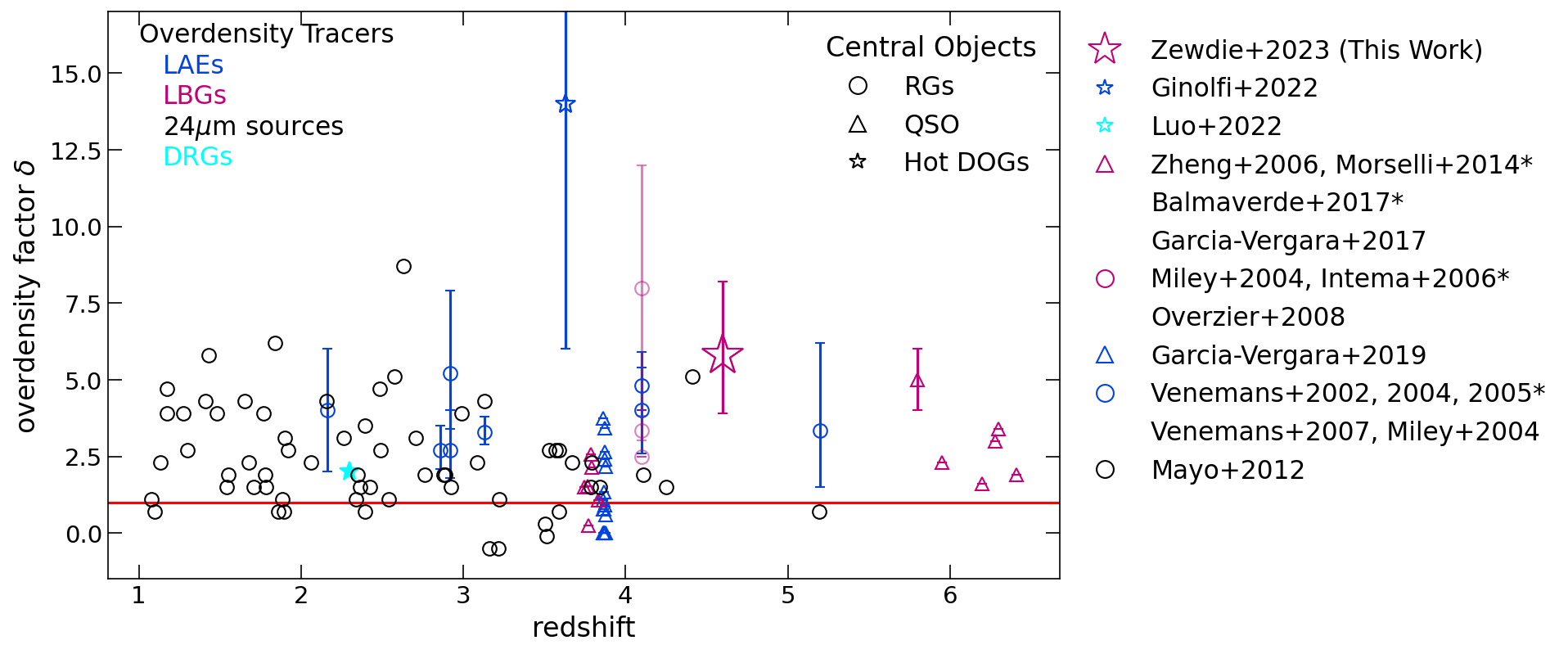}
\end{center}
\caption{The overdensity around high-redshift radio galaxies, quasars, and Hot DOGs as a function of redshift. We highlight the studies based on LBGs around quasars \citep[magenta triangles, ][]{2006Zheng, 2014Morselli, 2017Balmaverde, 2017GarcaVergara}, around radio galaxies \citep[magenta circles, ][]{2004Miley, 2006Intema, 2008Overzier} and Hot DOGs (magenta star, this work),  those based on LAEs  around quasars \citep[blue triangles, ][]{2019Garcia},  radio galaxies  \citep[blue circles, ][]{2002Venemans, 2004Venemans, 2005Venemans, 2007Venemans, 2004Miley}, and Hot DOGs \citep[blue star, ][]{2022Ginolfi} and a study based on DRGs around Hot DOGs \citep[cyan star, ][]{2022Luo}.  Overdensities of 24$\mu$m sources  around high-redshift radio galaxies are also shown \citep[gray dot, ][]{2012Mayo}. The red solid horizontal line indicates no overdensity. }
\label{dense}
\footnotesize $\ast$: These studies defined the overdensity instead as $\delta = N_{found}/N_{expected} - 1$. We transformed these estimates to reflect our definition of $\delta$ before adding them to the Figure. 
\end{figure*}

Studies based on a number of different techniques and wavelengths show that dense environments are not a unique property of Hot DOGs, but are shared by different types of luminous quasars, consistent with the evolutionary picture recently proposed by \cite{2022Assef} in which Hot DOGs correspond to a phase in the life of luminous quasars. Figure \ref{dense} shows a compilation of overdensities measured in quasar studies, which we detail below.

 \subsubsection{ LBGs Around Quasars and Radio Galaxies}
 
\cite{2017GarcaVergara}  studied LBGs around six quasars at $z\sim4$ using VLT/FORS narrow band imaging and found a strong LBG clustering with an overall overdensity of $\delta = 1.5$.  \cite{2014Morselli} also studied the overdensities of LBGs around four luminous z$\sim $6 quasars and found a significant galaxy overdensity of $\delta = 2.35$. This finding is also supported by \cite{2017Balmaverde} and \cite{2020Mignoli}, who obtained a similar result. \cite{2010Utsumi} and \cite{2006Overzier} independently investigated LBGs around quasars and found overdensities that exceeded the 99\% confidence level. \cite{2004Miley} studied the overdensity of LBGs around a radio galaxy at redshift $z = 4.1$ and found the environment to be 2.5 times denser than the field. Furthermore, when they restricted the inspected area to a 1 Mpc radius, they found that the field  of the radio galaxy exhibited an overdensity five times higher than the average blank field, which is a comparable result to our findings in similarly sized \textbf{regions}. \cite{2006Intema} conducted a study on the overdensity of LBGs in the vicinity of a radio galaxy and reported an overdensity of $\delta$=7$\pm$4. Similarly, \cite{2008Overzier} studied the overdensity of LBGs and LAEs around a radio galaxy and found an average overdensity of $\delta=3.53^{+0.47}_{-0.33}$. In addition, \cite{2018Ota} found high-density excess clumps of LBGs around quasars at $z=6.61$, and \cite{2023Kashino} and \cite{2023Matthee} conducted spectroscopic analysis and verified the existence of an overdensity of [OIII] emission surrounding the most luminous quasar at $z>6$.

\subsubsection{ LAEs Around Quasars and Radio Galaxies}
 
\cite{2019Garcia} probed the overdensity of Ly$\alpha$ emitters around 17  quasars at $z\sim 4$ using the VLT/FORS2 narrow band imaging and observed clustering with an average overdensity of $\delta=1.4\pm0.4$ although they note that 10 of the 17 targets have nominally underdense environments. They highlight the large cosmic variance inherent in quasar environments, with overdensities ranging from 0.0 to 3.75. \cite{2002Venemans} studied the environment of radio galaxies and found that the number density of Ly$\alpha$ emitters is $\delta=4\pm1.4$. They also evaluated the overdensity by taking into account the detected FWHM of the velocity distribution, which is four times smaller than the FWHM of the filter width, and revealed this radio galaxy field is 15 times denser than a blank field in Ly$\alpha$ emitters. \cite{2004Venemans} also found an overdensity of Ly$\alpha$ emitters of $\delta=3.35_{-1.85}^{+2.85}$ around radio galaxy at z=5.2.
 
 \subsubsection{ Mid-Infrared  Sources  Around Radio Galaxies}
 
\cite{2012Mayo} studied overdensities of 24 $\mu$m sources around 63 high-redshift radio galaxies between redshift $z=1-5.2$ using the using the Multiband Imaging Photometer for \textit{Spitzer} (MIPS) and found an average overdensity of $2.2\pm 1.2$ compared to the \textit{Spitzer} Wide-area InfraRed Extragalactic Survey (SWIRE) fields. In this large sample of high-redshift radio galaxies, they confirmed 11 protocluster candidates and identified 9 new. The selected protocluster candidates have an overdensity ranging from 3.1 to 8.7. Note that, in particular, \cite{2012Mayo} found a protocluster candidate of a high-redshift radio galaxy with an overdensity of $\delta$=5.1 at z=4.413. \cite{2013Wylezalek} studied the environment of obscured and unobscured radio-loud AGNs at $1.3< z< 3.2$ and found that 92\% of radio-loud AGNs are overdense, however, they did not quantify the overdensities using the delta parameter. As can be seen in Figure \ref{dense}, Hot DOGs sit towards the upper range of the diversity of environments around other types of luminous quasars, but mostly within it. This is consistent with Hot DOGs being one of the phases in the evolution of luminous quasars, as while the AGN properties are widely different between them, they live in qualitatively similar regions of the Universe.

 

 \section{Conclusions}\label{sec:Conclusions}
We present GMOS-S deep imaging observations of the Hot DOG, W2246$-$0526 in the \textit{r}-, \textit{i}-, and \textit{z}-bands, and we use them to identify potential LBG companions and characterize the density of its environment.  We use the \citetalias{2004VOuchi} and \citetalias{2006Yoshida} color selection criteria to identify LBG candidates, modified to account for the differences between the GMOS-S and Suprime-cam bands, assuming the using the  \cite{2003Shapley} LBG composite modified to include the additional IGM absorption expected at $z=4.601$.  Figure \ref{colordiagram} shows that both modified criteria can successfully isolate LBG candidates from foreground objects and brown dwarf stars (see Section \ref{modified} and Figure \ref{inter}). Using the modified \citetalias{2004VOuchi} and \citetalias{2006Yoshida} color selection, we found 37 and 55 LBG candidates (\textit{r}-dropouts), respectively, in an area of 23.7 arcmin$^2$ centered on W2246$-$0526.  We find the following main results:
 
 \begin{enumerate}
\item W2246$-$0526 lives in a dense environment as compared with the mean density of the Universe. Specifically, matching to \textit{z}-band magnitude depth of \citetalias{2004VOuchi} and \citetalias{2006Yoshida} studies, this corresponds to $\delta = 5.8^{+2.4}_{-1.9}$ times the surface density of LBGs expected in the field

\item  Figure \ref{reddein} shows that, in order to be selected, companion LBGs may need a combination of mild dust obscuration and excess IGM absorption over the typical expected amount. This suggests that we might be missing many potential companions to W2246$-$0526. In fact, we find that the \citetalias{2004VOuchi} selection does not identify either of the known, resolvable companions found by \cite{2016Tanio, 2018Tanio}, and the \citetalias{2006Yoshida} selection identifies only one of them.
   
 \item We estimated the UV luminosity function of the  selected LBG candidates around W2246$-$0526 in both modified selection criteria and found the presence of the LBG candidates in significantly higher space densities than in the studies of \citetalias{2004VOuchi} and \citetalias{2006Yoshida}, consistent with our findings based on the number densities. 
   
 \item We studied the density of LBG candidates as a function of distance from the Hot DOG, counting them in rings of 20 arcsec widths starting 2 arcsec away from W2246$-$0526. We did not observe a clear radial profile centered on the Hot DOG.  While W2246$-$0526 lives in an overdense environment, the overdensity is not clearly concentrated around the Hot DOG. This may be because a) the overdensity is in an early stage of collapse and has not yet virialized; b) although this Hot DOG is the most luminous object in the structure and radiates well above the Eddington limit, it is not the most massive or central galaxy in the structure; or c) the intense radiation emitted by the Hot DOG may affect the IGM transparency around it, deleting the radial pattern in the LBG overdensity.

\item  The W2246$-$0526 overdense environment is qualitatively consistent with statistical studies of {\tt{Spitzer}}/IRAC and submillimeter companions to Hot DOGs \citep{2015Assef, 2014Jones,  2017Fan} and with recent observational studies of the environment around the Hot DOGs W0410 at z=3.6  \citep{2022Ginolfi} and W1835 at z=2.35 \citep{2022Luo}.

\item Compared with radio galaxies and rest-frame UV bright quasars at a similar redshift, Hot DOGs live in somewhat denser environments (see Figure \ref{dense}). The higher overdensity suggests that Hot DOGs could be a good tracer for dense regions like proto-clusters at high redshift.

 \end{enumerate}
    Further studies of the environments of Hot DOGs, particularly at different redshifts and luminosities, and considering different ways of selecting the potential nearby counterparts, will push forward our understanding of SMBHs, galaxies, and large structure formation and evolution.



 \begin{acknowledgements}
We thank the referee for the constructive comments, which helped us to improve the paper. DZ gratefully acknowledges the support of ANID Fellowship grant No. 21211531.  RJA was supported by FONDECYT grant numbers 1191124 and 1231718. MA acknowledges support from FONDECYT grant 1211951 and ANID+PCI+REDES 190194. DZ, MA, RJA, and CM acknowledge support from ANID+PCI+INSTITUTO MAX PLANCK DE ASTRONOMIA MPG 190030. DZ, RJA, and MA were supported by the ANID BASAL project FB210003. C.-W. Tsai acknowledges support from the NSFC grant No. 11973051. H.D.J. was supported by the National Research Foundation of Korea (NRF) funded by the Ministry of Science and ICT (MSIT) of Korea (No. 2020R1A2C3011091, 2021M3F7A1084525, 2022R1C1C2013543). This research was partly carried out at the Jet Propulsion Laboratory, California Institute of Technology, under a contract with NASA. Based on observations obtained at the international Gemini Observatory, a program of NSF’s NOIRLab, which is managed by the Association of Universities for Research in Astronomy (AURA) under a cooperative agreement with the National Science Foundation on behalf of the Gemini Observatory partnership: the National Science Foundation (United States), National Research Council (Canada), Agencia Nacional de Investigaci\'{o}n y Desarrollo (Chile), Ministerio de Ciencia, Tecnolog\'{i}a e Innovaci\'{o}n (Argentina), Minist\'{e}rio da Ci\^{e}ncia, Tecnologia, Inova\c{c}\~{o}es e Comunica\c{c}\~{o}es (Brazil), and Korea Astronomy and Space Science Institute (Republic of Korea).


 \end{acknowledgements}

\begin{appendix} 
\section{LBG Candidates}
 The coordinates and magnitudes of all LBG candidates that were selected based on the modified \citetalias{2004VOuchi} and \citetalias{2006Yoshida} criteria are shown in Table \ref{LBGO}. Those selected solely by the \citetalias{2006Yoshida} criteria are shown in Table \ref{LBGY}. We also list the previously detected sources by \cite{2016Tanio} and \cite{2018Tanio} in Table \ref{CT}. Stamps of 8$\arcsec\times 8\arcsec$ are shown in Figures  \ref{stamps} and \ref{stampy} for all LBG candidates. The stretch in the stamps of all bands is linear and normalized to the \textit{i}-band flux of each source. We also show the stamps of the detected sources by \cite{2016Tanio} and \cite{2018Tanio} using the deep ALMA observations and stamps of W2246-0526 in Figure  \ref{stampt}.


	\begin{table*}
			\caption{The list of selected LBG candidates in order of \textit{i}-band magnitude from brightest to faintest using the modified selection criteria of \citetalias{2004VOuchi} and \citetalias{2006Yoshida}, the magnitudes are measured in 2$\arcsec$ apertures.}
		\label{LBGO}
  $$
	\begin{array}{c c c c c c c c c}		
			\hline
		\noalign{\smallskip}
			\text{LBG candidates}  & Ra    & Dec &     r_{mag}   & r_{magerr} & i_{mag}  &  i_{magerr} & z_{mag} &  z_{magerr} \\ 
				\noalign{\smallskip}
			\hline
			\noalign{\smallskip} 
LBGC1   & 341.5000812  & -5.4621209 & 25.91  &  0.06  &   24.40   &   0.02   &          24.21 &0.04\\            
LBGC2   & 341.5238785  & -5.4738352 & 26.62  &  0.11  &   24.98   &   0.04   &          24.95 &0.08\\            
LBGC3   & 341.5634960  & -5.4611731 & 26.93  &  0.15  &   25.07   &   0.05   &          24.60 &0.05\\            
LBGC4   & 341.5406383  & -5.4077323 & 26.98  &  0.16  &   25.18   &   0.06   &          25.55 &0.14\\            
LBGC5   & 341.5674046  & -5.4257313 & 26.93  &  0.15  &   25.45   &   0.07   &          25.33 &0.11\\            
LBGC6   & 341.5546196  & -5.4564413 & 27.22  &  0.20  &   25.51   &   0.08   &          25.84 &0.18\\            
LBGC7   & 341.5676332  & -5.4255572 & 26.99  &  0.16  &   25.54   &   0.08   &          25.92 &0.19\\            
LBGC8   & 341.5167494  & -5.4631821 & 27.74  &  0.32  &   25.55   &   0.08   &          24.91 &0.07\\            
LBGC9   & 341.5434564  & -5.4698341 & 27.06  &  0.17  &   25.61   &   0.08   &          25.47 &0.13\\            
LBGC10  & 341.5683588  & -5.4263881 & 28.22  &  0.51  &   25.64   &   0.09   &          25.39 &0.12\\            
LBGC11  & 341.4862233  & -5.4518122 & 27.37  &  0.23  &   25.72   &   0.09   &          25.55 &0.13\\            
LBGC12  & 341.4896542  & -5.4596500 & 27.46  &  0.25  &   25.86   &   0.11   &          25.71 &0.16\\            
LBGC13  & 341.5260347  & -5.4651164 & 27.48  &  0.25  &   25.88   &   0.11   &          25.66 &0.15\\            
LBGC14  & 341.5436721  & -5.4741600 & 27.53  &  0.27  &   25.95   &   0.12   &          26.52 &0.34\\            
LBGC15  & 341.5239417  & -5.4254533 & 28.58  &  0.71  &   26.01   &   0.12   &          26.01 &0.21\\            
LBGC16  & 341.5683885  & -5.4150181 & >29.0  &  ...   &   26.01   &   0.12   &          25.73 &0.16\\            
LBGC17  & 341.4860620  & -5.4519042 & 28.53  &  0.68  &   26.03   &   0.13   &          25.56 &0.14\\            
LBGC18  & 341.5175633  & -5.3935374 & 28.14  &  0.47  &   26.09   &   0.13   &          25.98 &0.20\\            
LBGC19  & 341.5725185  & -5.4205112 & 28.15  &  0.48  &   26.10   &   0.13   &          25.67 &0.15\\            
LBGC20  & 341.5289983  & -5.4473865 & 27.79  &  0.34  &   26.16   &   0.14   &          26.40 &0.30\\            
LBGC21  & 341.5316215  & -5.4170330 & 27.91  &  0.38  &   26.17   &   0.14   &          25.73 &0.16\\            
LBGC22  & 341.4826847  & -5.4571690 & >29.0  &  ...   &   26.20   &   0.15   &          25.69 &0.15\\            
LBGC23  & 341.5347913  & -5.4452524 & >29.0  &  ...   &   26.24   &   0.15   &          26.44 &0.31\\            
LBGC24  & 341.5055473  & -5.4173291 & >29.0  &  ...   &   26.24   &   0.15   &          26.00 &0.21\\            
LBGC25  & 341.5244762  & -5.4688419 & 27.76  &  0.33  &   26.30   &   0.16   &          26.38 &0.30\\            
LBGC26  & 341.5619358  & -5.4686734 & 28.11  &  0.46  &   26.35   &   0.17   &          26.52 &0.34\\            
LBGC27  & 341.5160667  & -5.4269681 & 27.95  &  0.40  &   26.36   &   0.17   &          26.06 &0.22\\            
LBGC28  & 341.5059390  & -5.4178372 & >29.0  &  ...   &   26.41   &   0.18   &          26.53 &0.34\\            
LBGC29  & 341.5148175  & -5.4221763 & >29.0  &  ...   &   26.41   &   0.18   &          26.28 &0.27\\            
LBGC30  & 341.5530543  & -5.4439086 & 28.37  &  0.58  &   26.44   &   0.19   &          25.94 &0.20\\            
LBGC31  & 341.5428250  & -5.4423416 & >29.0  &  ...   &   26.48   &   0.19   &          25.86 &0.18\\            
LBGC32  & 341.5230371  & -5.4030845 & 28.54  &  0.69  &   26.59   &   0.21   &          26.04 &0.21\\            
LBGC33  & 341.5228025  & -5.4275428 & 28.48  &  0.65  &   26.63   &   0.22   &          26.21 &0.25\\            
LBGC34  & 341.5267696  & -5.4423843 & 28.73  &  0.82  &   26.64   &   0.22   &          26.07 &0.22\\            
LBGC35  & 341.5189382  & -5.4619869 & >29.0  & ...    &   26.65   &   0.23   &          26.40 &0.30\\            
LBGC36  & 341.4958418  & -5.4524641 & >29.0  & ...    &   26.72   &   0.24   &          26.45 &0.32\\            
LBGC37  & 341.5494098  & -5.4744106 & >29.0  &  ...   &   27.04   &   0.33   &          26.46 &0.32\\            
 
		\noalign{\smallskip}
	\hline
\end{array}
$$
	\end{table*}

	\begin{table*}
			\caption{The list of selected LBG candidates in order of \textit{i}-band magnitude from brightest to faintest using the modified selection criteria of \citetalias{2006Yoshida}}
		\label{LBGY}
		$$ 
	\begin{array}{c c c c c c c c c}		
		\hline
		\noalign{\smallskip}
		\text{LBG candidates}  & Ra    & Dec &     r_{mag}   & r_{magerr} & i_{mag}  &  i_{magerr} & z_{mag} &  z_{magerr} \\ 
		\noalign{\smallskip}
		\hline
		\noalign{\smallskip} 
LBGC38& 341.5315566 & -5.4424653 & 25.84  & 0.05    & 24.48      &  0.03    &24.43 & 0.05\\           
LBGC39& 341.5348617 & -5.4496508 & 26.63  & 0.11    & 25.27      &  0.06    &25.73 & 0.16\\           
LBGC40& 341.5879426 & -5.4218498 & 26.69  & 0.12    & 25.27      &  0.06    &25.32 & 0.11\\           
LBGC41& 341.5128808 & -5.4353647 & 26.64  & 0.12    & 25.28      &  0.06    &25.23 & 0.10\\           
LBGC42& 341.5497232 & -5.4772553 & 26.62  & 0.11    & 25.28      &  0.06    &25.33 & 0.11\\           
LBGC43& 341.5290441 & -5.4836258 & 26.58  & 0.11    & 25.31      &  0.06    &25.47 & 0.13\\           
LBGC44& 341.5136125 & -5.4468992 & 26.76  & 0.13    & 25.35      &  0.07    &25.54 & 0.13\\           
LBGC45& 341.5174896 & -5.4302933 & 26.82  & 0.14    & 25.47      &  0.07    &25.44 & 0.12\\           
LBGC46& 341.5306453 & -5.4806720 & 26.77  & 0.13    & 25.49      &  0.07    &26.27 & 0.27\\           
LBGC47& 341.5376427 & -5.4820949 & 26.96  & 0.16    & 25.50      &  0.08    &25.26 & 0.10\\           
LBGC48& 341.5623409 & -5.4306788 & 26.94  & 0.15    & 25.68      &  0.09    &25.80 & 0.17\\           
LBGC49& 341.5373545 & -5.4108363 & 27.16  & 0.19    & 25.89      &  0.11    &26.14 & 0.24\\           
LBGC50& 341.5324556 & -5.4005839 & 27.44  & 0.24    & 26.16      &  0.14    &26.44 & 0.31\\           
LBGC51& 341.5418400 & -5.4629014 & 27.43  & 0.24    & 26.19      &  0.15    &26.35 & 0.29\\           
LBGC52& 341.5611256 & -5.4731423 & 27.65  & 0.30    & 26.22      &  0.15    &26.40 & 0.30\\           
LBGC53& 341.5673180 & -5.4646407 & 27.61  & 0.29    & 26.34      &  0.17    &26.34 & 0.28\\           
LBGC54& 341.5493831 & -5.4784556 & 27.85  & 0.36    & 26.46      &  0.19    &26.34 & 0.29\\           
LBGC55& 341.5210883 & -5.4012651 & 28.34  & 0.57    & 26.79      &  0.26    &26.46 & 0.32\\           
 
	\noalign{\smallskip}
		\hline
	\end{array}
$$ 
	\end{table*}

	\begin{table*}
			\caption{\cite{2016Tanio} and \cite{2018Tanio} list of two spectroscopically confirmed and potential companions in order of \textit{i}-band magnitude from brightest to faintest found through deep ALMA observations.}
		\label{CT}
		$$ 
	\begin{array}{c c c c c c c c c}		
		\hline
		\noalign{\smallskip}
		\text{LBG candidates}  & Ra    & Dec &     r_{mag}   & r_{magerr} & i_{mag}  &  i_{magerr} & z_{mag} &  z_{magerr} \\ 
		\noalign{\smallskip}
		\hline
		\noalign{\smallskip} 
        U1& 341.5319518 & -5.4427905 & 24.02 & 0.01  &   23.41  & 0.01 & 22.87 & 0.01\\ 
        C3 & 341.5315566 & -5.4424653 & 25.84 & 0.05  &   24.48  & 0.03 & 24.43 & 0.05\\
        C2 & 341.5326256 & -5.4440949 & 25.51 & 0.04  &   24.67  & 0.03 & 24.52 & 0.05\\                        
        Knot& 341.5322715 & -5.4439102 & 25.13 & 0.03  &   24.92  & 0.04 & 24.58 & 0.05\\  
        U3& 341.5300995 & -5.4454121 & 25.77 & 0.05  &   25.42  & 0.07 & 24.58 & 0.09\\   
        U2& 341.5303579 & -5.4416845 & 26.64 & 0.11  &   25.59  & 0.08 & 25.52 & 0.13\\

	\noalign{\smallskip}
		\hline
	\end{array}
$$ 
	\end{table*}

\begin{figure*}[ht]
\begin{center}
\includegraphics[scale=0.53]{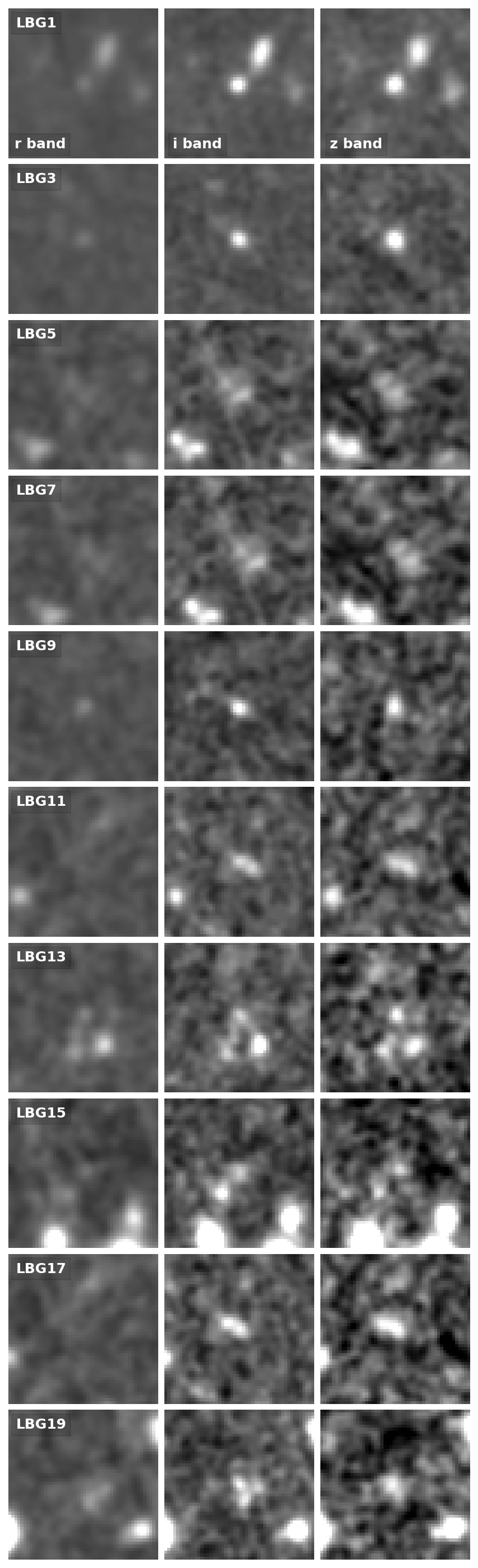} \hspace{0.1cm} \includegraphics[scale=0.53]{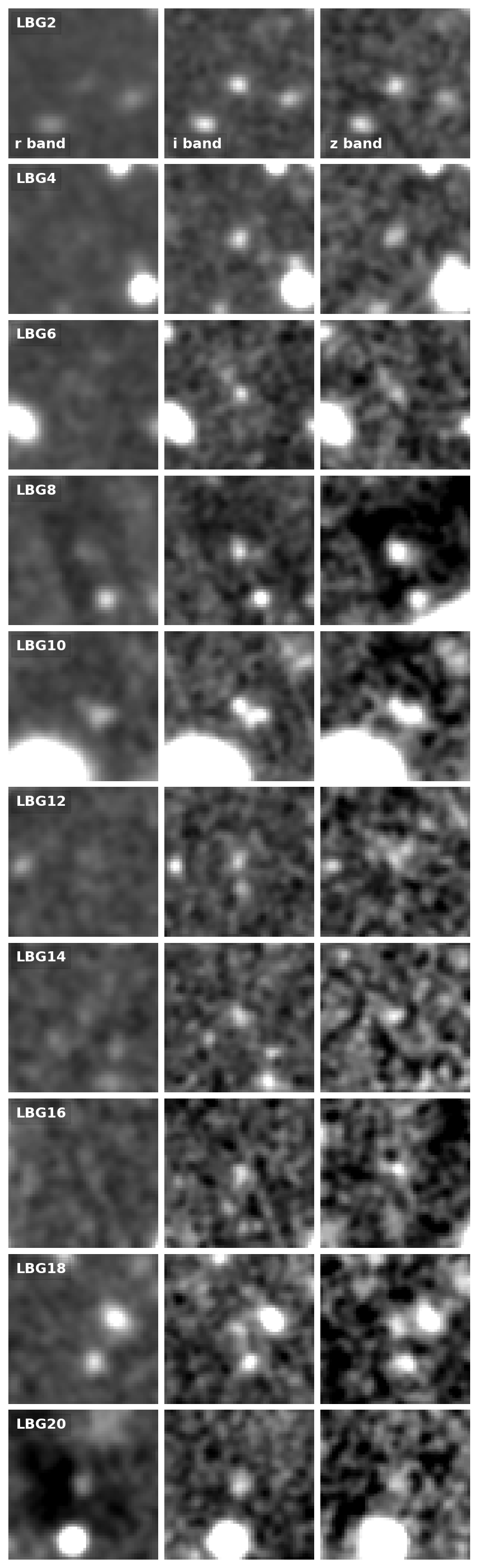}\\
\end{center}
\end{figure*}

\begin{figure*}[ht]
\begin{center}
\includegraphics[scale=0.58]{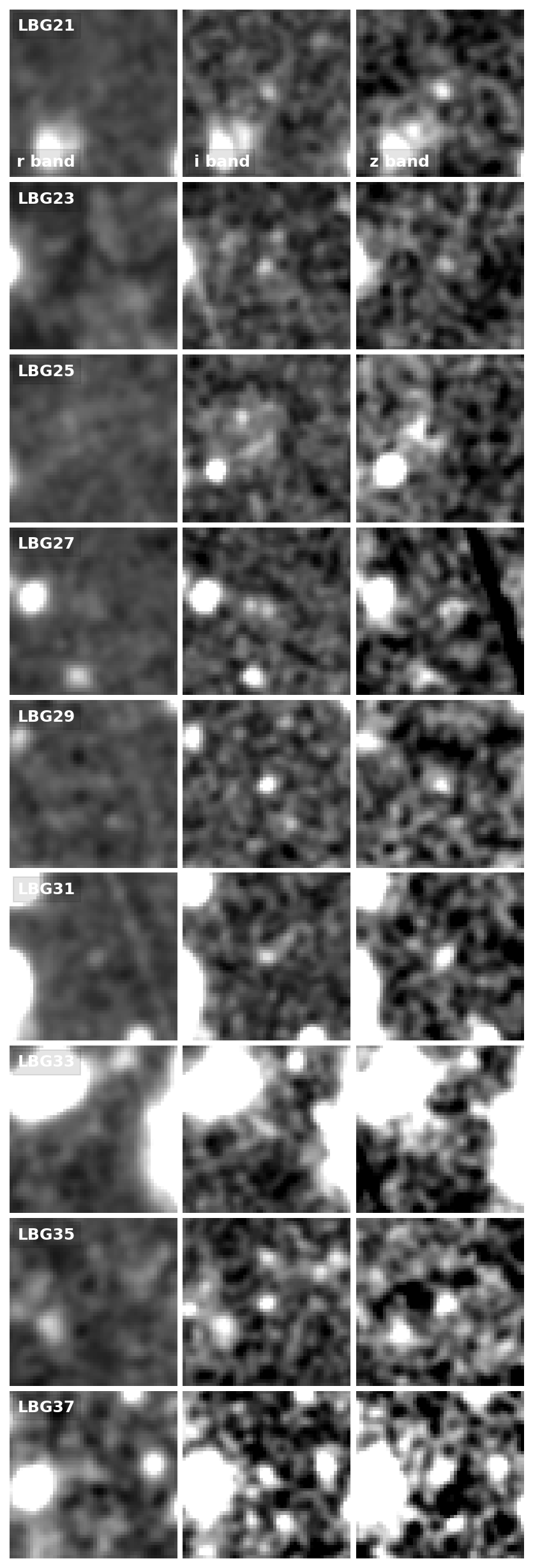} \hspace{0.1cm} \includegraphics[scale=0.58]{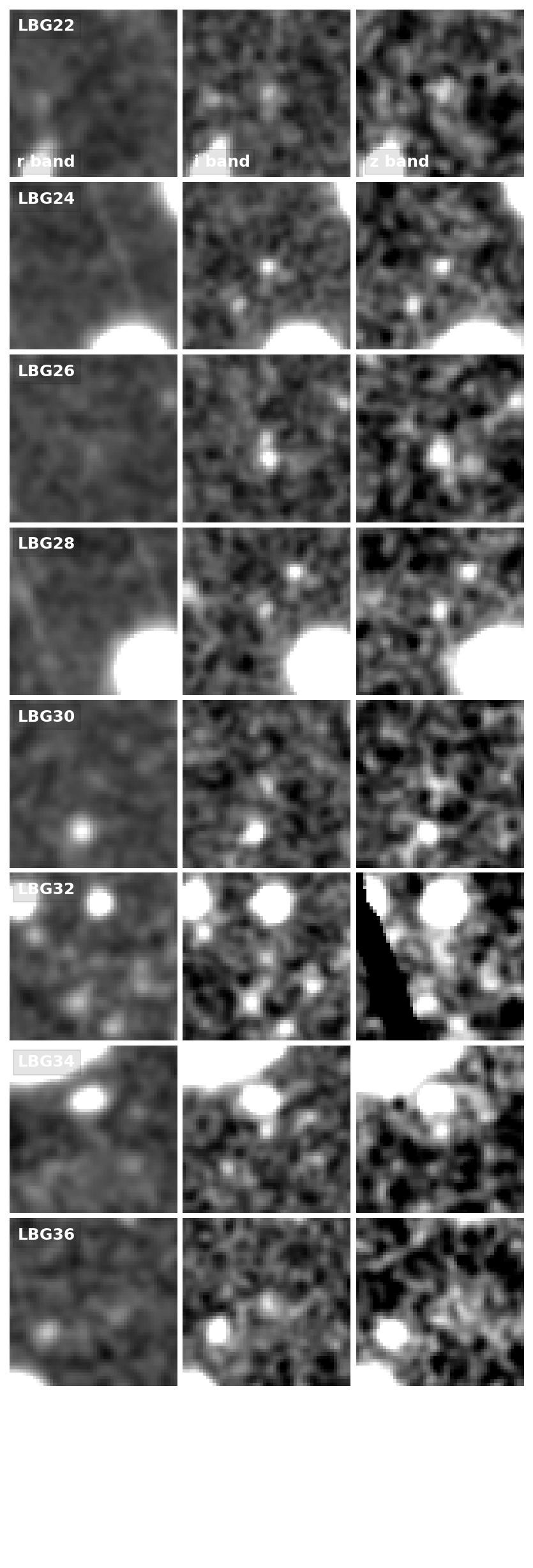}\\
\end{center}
\caption{Postage stamps (8$\arcsec \times 8\arcsec$) of the selected LBG candidates using the modified criteria of \citetalias{2004VOuchi} and \citetalias{2006Yoshida} in each band.}
\label{stamps}
\end{figure*}

\begin{figure*}[ht]
	\begin{center}
		 \includegraphics[scale=0.58]{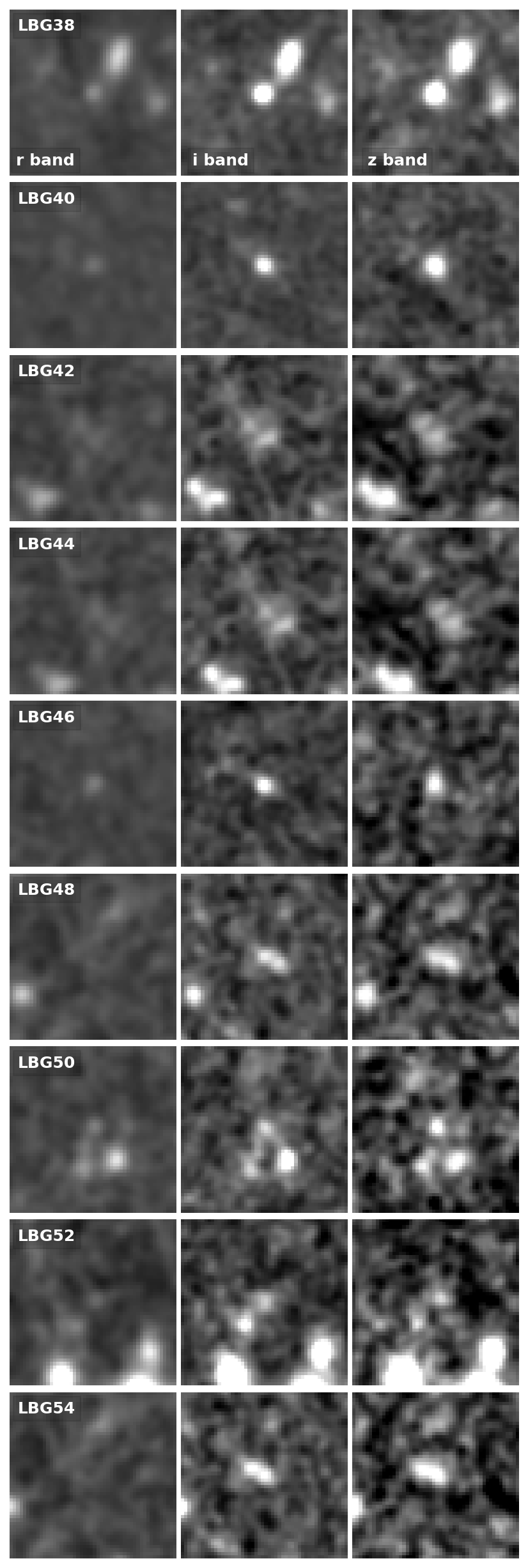}  \hspace{0.1cm} \includegraphics[scale=0.58]{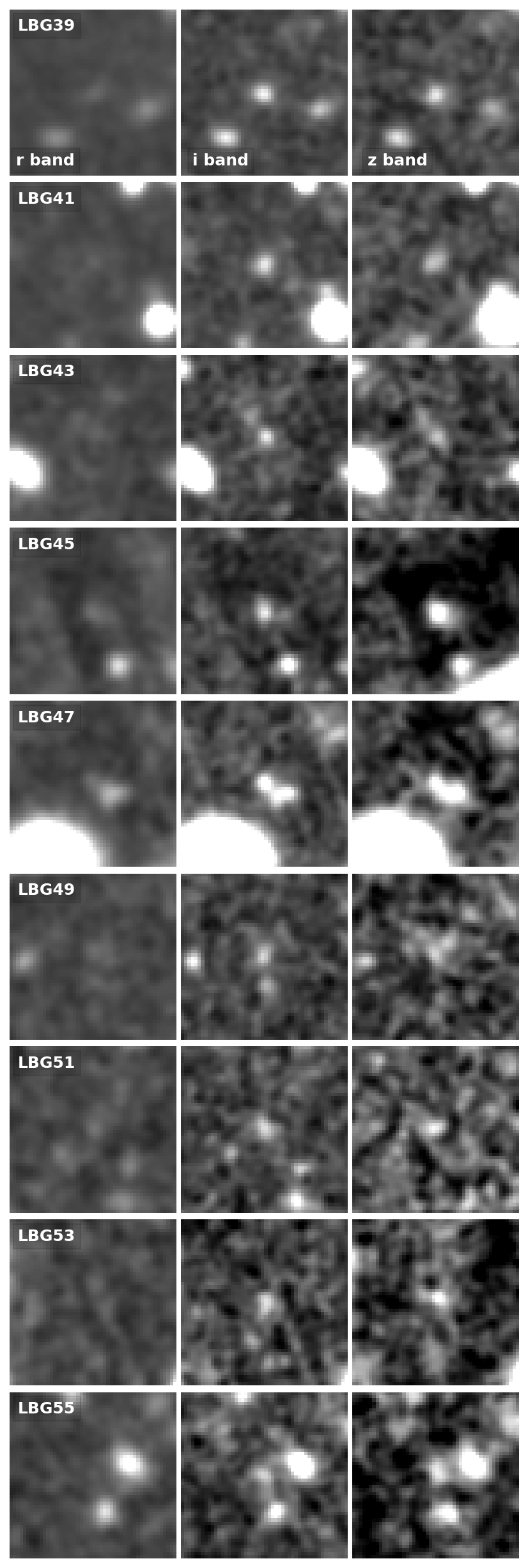} \\
  
	\end{center}
	\caption{ Postage stamps (8$\arcsec \times 8\arcsec$) of the 18 LBG candidates that were selected only by the modified criteria of \citetalias{2006Yoshida}.}
	\label{stampy}
\end{figure*}

\begin{figure*}[ht]
	\begin{center}
		\includegraphics[scale=0.68]{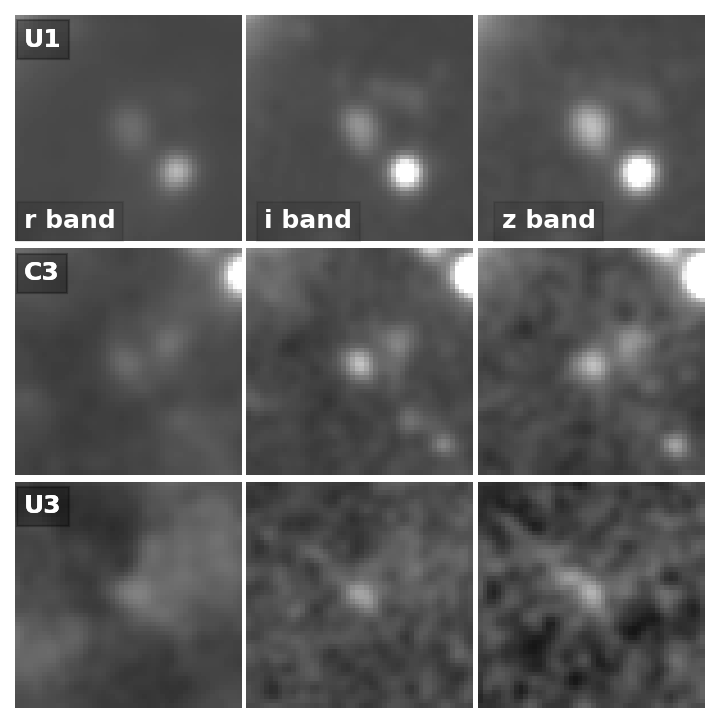} \hspace{0.1cm} \includegraphics[scale=0.68]{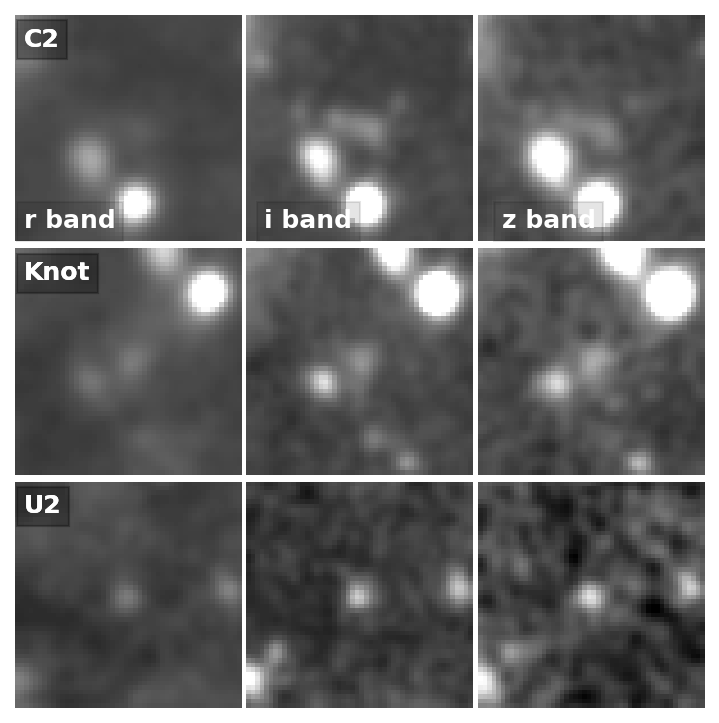}
   \includegraphics[scale=0.33]{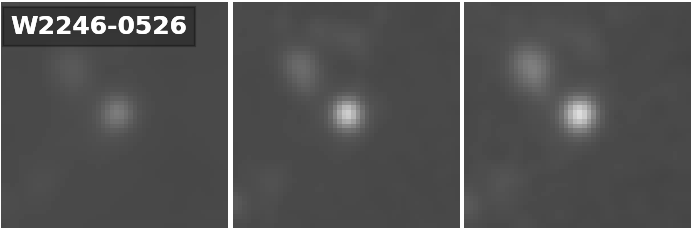}  \hspace{0.1cm} \includegraphics[scale=0.34]{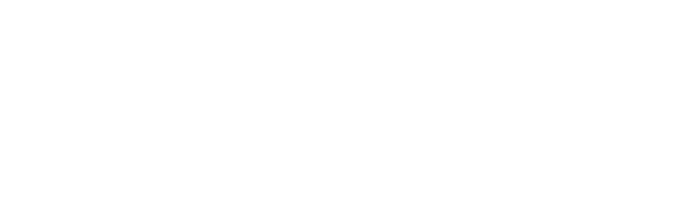}
	\end{center}
	\caption{Postage stamps (8$\arcsec \times 8\arcsec$) of the W2246$-$0526, 2 spectroscopically confirmed companions (C2 and C3) and 4 potential companions (U1, U2, U3, and knot) that were detected using deep ALMA observation by \cite{2016Tanio} and \cite{2018Tanio}. All those companions with W2246$-$0526 are shown in Figure \ref{stampw} }
	\label{stampt}
\end{figure*}

 \end{appendix}

\end{document}